\documentstyle[12pt,aasms4]{article}
\def\lapprox{\hbox{\lower .8ex\hbox{$\,\buildrel < \over\sim\,$}}}
\def\gapprox{\hbox{\lower .8ex\hbox{$\,\buildrel > \over\sim\,$}}}

\begin{document}

\bigskip
\bigskip
\bigskip
\bigskip

\title
{The cosmic gamma-ray background in the MeV range}

\author
{P. Ruiz--Lapuente \altaffilmark{1,2}, M. Cass\'e \altaffilmark{3,4}
 \& E. Vangioni--Flam \altaffilmark{3}}

\altaffiltext
{1}{ Department of Astronomy, University of Barcelona, Mart\'\i\ i Franqu\'es
1, E--08028 Barcelona, Spain. E--mail: pilar@mizar.am.ub.es}

\altaffiltext
{2}{Max--Planck--Institut f\"ur Astrophysik, Karl--Schwarzschild--Strasse 1,
D--85740 Garching, Federal Republic of Germany. E--mail:
pilar@MPA--Garching.MPG.DE}

\altaffiltext
{3} {Institut d'Astrophysique de Paris, 98bis Bd Arago, 75014 Paris, 
     France.  E-mail : flam@iap.fr, casse@iap.fr}

\altaffiltext
{4} {Service d'Astrophysique, 
DSM/DAPNIA/CEA, Orme des Merisiers, Gif 
         sur Yvette, Cedex, France}

\slugcomment{{\it Running title:} $\gamma$--ray background}

\begin{abstract}

 The $\gamma$--ray background from supernovae is calculated 
 on the basis of the reconstructed efficiency of supernova
 explosions from star formation at various redshifts. 
 The calculations presented here show how Type Ia SN rates 
 compatible with the results of optical supernova
 searches give a background emission in the MeV range
 that can explain the extragalactic emission  
 measured by COMPTEL and SMM. Star formation 
 histories reaching maximum values 
 of 0.3--0.4 M$_{\odot}$ yr$^{-1}$ Mpc$^{-3}$
 at z$\sim$ 1.5 (and/or possibly keeping that value to higher z) 
 give a $\gamma$--ray background compatible
 with observations while they also predict 
 the observed supernova rates per comoving volume. 
 While supernova rates are sensitive to the
 cosmology and details of star formation history along z, 
 the $\gamma$--ray background in the MeV range 
 is very weakly dependent on the cosmology and compatible 
 with a range of star formation histories. It is mostly sensitive
 to the star formation rate at z$\sim$ 1--2, but the activity in
 forming stars at even higher z has an influence as well on 
 the absolute level of the 
 cosmic $\gamma$--ray background: the time elapsed by the SNe Ia
 progenitors until they explode has a broad distribution 
 and the rates of SNe Ia 
 at z$\sim$ 1.5--2 depend on the formation frequencies of their 
 progenitor stars at earlier epochs.
 The mutual 
 consistency of predictions of optical rates and 
 integrated emission in the $\gamma$--ray domain place in
 a firm ground the Type Ia supernova identification
 as the astrophysical source responsible for this cosmic background.

\end{abstract}

\keywords{cosmology: diffuse radiation ---stars: supernovae: general 
---gamma--rays: theory}

\section{Introduction}

    All--sky surveys at various wavelengths reveal the presence of 
   a background of diffuse emission whose origin is diverse and
   appears unclear at some specific energies.  
   In the X--ray domain, it has been recognized that
   AGNs and Seyfert galaxies provide most of the
   emission (Madau, Ghisellini \& Fabian 1994; 
   Comastri et al. 1995). That emission covers from a few keV till several
   hundred keV. At very high energies, in the hundred MeV till 
   TeV range blazars seem to be responsible for the observed
   fluxes (Zdziarski 1996; Sreekumar et al. 1998).

    The measurements provided by the APOLLO 15/16 missions in the 
   MeV range suggested a possible change in the
   energy slope of the background towards a flatter spectrum
   in the 1 to 5 MeV region
   (Trombka et al. 1977). The significance of that excess
   over a smooth power--law connection of low and high--energy data 
   was only 1.5 $\sigma$. However, at just that energy range 1--5 MeV
    AGNs and blazars were known  not to emit particularly strongly. 
   Since those first empirical results came out, it became relevant
    to identify the 
   astrophysical sources adding in that energy interval to
   the diffuse extragalactic background because
   in the lack of astrophysically ordinary contributions to that
   emission, other more exotic considerations would be raised -i.e. 
   for some dark matter candidates annihilation with other particles 
   or decay might occur and give a significant integrated signal 
   in the gamma--ray domain (see, for instance, Kamionkowski 1995; Ellis 2000).
   Some of the theoretical candidates to explain that MeV emission 
   were already proposed, but
   the final comparison with the data had to  wait till the reanalisis 
   of the Apollo data and the measurements done
   by HEAO-A4 (Kinzer et al. 1997), the Solar Maximum Mission
   (Watanabe et al. 1999a), and more recently the availability
   of the diffuse $\gamma$--ray background measurements by COMPTEL 
   (Kappadath et al. 1996; Weidenspointer 2000).

   The measurement of the  extragalactic $\gamma$--ray background 
   in its present stage has, therefore, taken a great deal of 
   training in the understanding of the response
   of the instruments on board of the different observatories . 
   The analysis of HEAO--A4, SMM and COMPTEL data has just
   released a consistent picture where the slope of the emission in the hard
   X-ray domain decreases with increasing energy 
    from a few hundred keV range and meets a softer
   shape at about 
   10 MeV implying
   smaller fluxes than suggested by the early data, but 
   revealing the need of an intense extragalactic source 
   in that MeV window.

   Already, in the late sixties, Clayton \& Silk (1969) had suggested 
   that Type Ia supernovae (SNe Ia) would give rise 
   to a diffuse $\gamma$--ray
   background in the MeV range. The level of that background emission
   is discusssed in various works (The et al. 1993; Watanabe et al. 1999b). 
   In the work of Watanabe et al. (1999b) supernovae are seen to
   provide considerable emission, but the flux level, according to 
   these authors, could be too low as compared with the observed
   measured level. 
   Alternatively, there have been attempts to attribute the origin of
   that diffuse background to blazars. As the known population
   of blazars does not emit strongly enough
   in the MeV regime (McNaron--Brown et al. 1995), in order 
   to explain the observations, speculation has arisen on 
   a new population of blazars,
   i.e., ``MeV blazars'' (Bloemen et al. 1995; Blom et al. 1995a,b). 
   As another possibility to explain this 
     $\gamma$--ray flux,
    Stecker et al. (1999) have 
    come up with the suggestion that the nonthermal electron tail
    of Seyfert galaxies would be responsible for originating the
    MeV emision by hardening the soft X--ray photons arising from
    the accretion of material around the galactic black hole. 
    Even stronger constraints have been raised on the possibility
    that the known population of blazars would account for the gamma--ray
    background in an energy range above the MeV (for 
    some recent alternatives to the blazar hypothesis see Loeb and Waxman 2000;
    Dar \& de Rujula 2000).

     In relation to the background predictions, the element which 
    introduces the largest uncertainty 
    for any suggested source is the number of events
    up to high redshift -i.e. rate or luminosity function.
    In the case of supernovae, as it will be shown 
    here, it is possible to give a precise answer on their 
    contribution by comparing 
    the gamma--ray observations with the number of SNe Ia 
    exploding up to high redshifts.  
    We are starting to have a clear idea 
    of the efficiency  of stars in producing SNe Ia up to high z, and 
    we can make predictions for the cosmic $\gamma$--ray background 
    with good reliability.

\section{The star formation rate and supernovae }

    The number of supernova explosions at high z is a direct function of the 
   star formation history and of the efficiency in producing supernovae
   along cosmic time. For the $\gamma$-ray background, we are
   mostly concerned with the efficiency in getting supernovae from 
   low and intermediate--mass stars in binary systems, i.e., Type Ia
   supernovae.  Type II supernovae
   coming from massive stars do not provide significant
   emission in the $\gamma$--ray domain. 
   
    Much has been learnt recently on the evolution of the star
   formation process. 
   The original attempt from Madau et al. (1996) of deriving 
   star formation rates from the UV emission density at high z 
   in the Hubble Deep Field revealed an increasing star formation towards
   higher z. The star formation history reconstructed by Madau, Pozzetti 
   \& Dickinson (1998)  
   peaks at z$\sim$ 1.5 with a range of
   0.12--0.17 M$_{\odot}$ yr$^{-1}$ Mpc$^{-3}$ 
   to fall again at higher z. Soon that decrease at z $ >$ 1.5  was 
   questioned by other different approach: the derivation of the
   star formation history from Lyman break galaxies (Steidel et al. 1998).
   Steidel et al. (1998) found that the spectroscopic properties of 
   the galaxy samples at z $\sim$ 3 and z $\sim$ 4 are indistinguishable,
   as are the luminosity function shapes and the total UV luminosity 
   density between z $\sim$ 3 and $\sim$ 4. From that work it is suggested
   that the star formation rate does not decrease at z $>$ 2 but levels 
   off. That conclusion of a star formation rate at z $>$ 2 higher than 
   the one obtained by Madau et al. (1996) is also supported from results
   at long wavelengths. The study of star formation with the
   submillimeter SCUBA array by Hughes et al. (1998) reveals significant 
   dust enshrouded star formation at high z.  The star formation rate
   density over the range  2 $ <$ z $ <$ 4 would be at least five
   times higher 
   than that inferred from the UV emission of the HDF galaxies. 
   Blain et al. (1998) reconstruct the star formation histories 
   compatible with the observations at various wavelengths incorporating
   the SCUBA results and the derivations from chemical evolution at high 
   z. Some of the compatible star formation histories peak at a z closer to
   2 (rather than 1 or 1.5).
   Figure 1 shows the star formation histories that are going
   to be explored in this work. In general, preference from several approaches
   is being buildt up for a peak at z $\sim$ 1--1.5 (Madau et al. 1998;
   Steidel et al. 1998; Blain et al. 1998). It has been pointed out against a 
   peak in the star formation at z $\sim$ 2 (Hughes et al. 1998), that 
   the submillimeter results
   have the advantage
   of unveiling star formation enshrouded by dust, but they suffer from the
   lack of real redshift identification of the observed emission. 
  
   Further evidence on star formation pointing to a larger star formation
   rate at high z than that inferred from UV--optical observations
   comes from measurements of the extragalactic background light (EBL) 
   by ISO (Rowan--Robinson et al. 1997; Flores et al. 1998; 
   Elbaz et al. 1999) and the 
   FIRAS and DIRBE experiments on board of COBE (Dwek et al. 1998; Fixsen
   et al. 1998). 
   The COBE results suggest at least a SFR twice as high at z $\sim$ 1 
   than that determined by Madau et al. (1998).

    There are as well open issues on how
  the star formation history evolves  between the local value and
  that at z close to 1. The $\gamma$--ray emission of 
  astrophysical sources from z $\sim$ 0 to z $\sim$ 2 will be indeed more
  relevant than the emission at z beyond 2 for setting the absolute
  flux of the background observed. 
    The strong dilution factor 
    linked to the square of the distance and the cosmological factors
    1+z entering in the flux observed  at earth from sources at very
    high z, add up to make the emission at very high z irrelevant 
    for our detection capabilities as 
    compared with the flux at z lower than 2. Most of 
   the $\gamma$--ray photons 
    received should come from supernovae that explode at
    z $\le$ 1.5. Their progeniitor systems were formed, however, at a
   much larger z. This makes relevant as well for the calculations
   the level of the SFR at z $>$ 2. 
   Determination of how much flux is received from
   SNeIa is grounded here in observations. 
  In the next paragraph, 
  we address the efficiency
  of SNeIa production in relation to the star formation rate determinations
  from the local universe to z$\sim$ 1.

\section{Efficiency of SNe Ia production}

     In parallel to the studies of star formation at high z, the 
     research on how many SNe Ia result from star formation
     activity at high z has now made a step forward. Since the first
    supernova  measurement at high z by the Supernova Cosmology Project (SCP)
   (Pain et al. 1996), searches of 
     SNe Ia centered at various z have provided the evolution
     of that rate up to z $\sim$ 0.55
     (Hamuy \& Pinto 1999; Hardin et al. 1998; Pain et al. 1999).
     Results are summarized in Table 1.

     As it has been pointed out (Ruiz--Lapuente, Burkert \& Canal
     1995, 1997; Madau, Della Vella \& Panagia 1998; Sadat et al. 
     1998; Dahlen \& Fransson 1998; 
     Yungelson \& Livio 1998, 1999), the better knowledge 
     of the SNe Ia rates at all redshifts allow us to identify
     the stellar systems leading to Type Ia explosions.

     Here we combine the knowledge of the SNe Ia rates
     at all redshifts 
     with that of the average star formation rate (discussed in the 
     previous section) in order to investigate whether the 
     efficiency in producing 
     Type Ia SNe has changed along cosmic time. Such numbers, when 
     compared with theoretical predictions, not only shed light on 
     the supernova progenitor issue but allow us to give a 
     clear  answer on the level of the cosmic $\gamma$-ray background 
     provided by supernovae of Type Ia. 
    
     A very useful quantity to introduce is the ``efficiency''
      in producing SNe Ia out of star formation, defined as
      the rate of SNe Ia
     at a given z per comoving volume (SN yr$^{-1}$ Mpc$^{-3}$) 
     divided by the star formation rate at the same z
     (in M$_{\sun }$ yr$^{-1}$ Mpc$^{-3}$). This value is independent
     of the cosmology assumed and refers to the number of SNe Ia per
     unit mass that has gone into star formation.

   $${\cal E}_{SNeIa}(z) = \Re_{Ia}(z)\ yr^{-1}\ 
 Mpc^{-3}\ /\ \dot \rho_{*}(z)\ 
   M_{\odot}\ yr^{-1}\ Mpc^{-3}\eqno(1)$$

     Any change in this efficiency for producing
     SNe Ia tells us about the timescale of the explosion as well 
     as about the evolutionary effects in the production of Type Ia
     supernovae. The effect of the delay between star formation and
     explosion is illustrated in Figure 2. A short timescale 
     of the order to $\sim$ 10$^{8}$ yr gives rise to a 
     constant ``efficiency'' along z, whereas
     a timescale of  a few 10$^{9}$ yr produces
     an accumulation of events towards low z. Suggested 
     metallicity effects reducing the number of
     type Ia explosions at z $>$ 1
    (Kobayashi et al. 1998) would also be reflected in the
     efficiency mentioned above by a  drop in that
     quantity. 

\bigskip

    Little direct evidence is so far available on the nature of
    the binary system which is responsible for Type Ia SN
    explosions. We know that the exploding star is a 
     carbon--oxygen white dwarf (C+O WD)
    and that there is no compact object left by the explosion. 
    Stellar evolution arguments tell
    us as well  that those explosions take place 
    in binary systems. But so far, it has not been possible to  
    point clearly to a unique type of binary system as responsible for 
    the Type Ia phenomenon
    (Ruiz--Lapuente, Canal \& Isern 1997 
    and references therein).  
    Two candidate systems seem favored. One of
    them is a pair of C+O WDs in a binary system that merge
    as their orbit shrinks due to the emission of 
    gravitational wave radiation (Iben \& Tutukov 1984; Webbink
    1984). This system is
    refered as to {\it double degenerate system}.
    The other one is
    a WD which accretes material from a Roche--lobe 
    overfilling non-WD companion 
    (WD plus Roche--lobe filling subgiant or
    giant). The system looks as an Algol-type
    binary. It contains only a WD and therefore is a
    {\it single--degenerate system} (Whelan \& Iben 1973;
     Branch et al. 1995; Hachisu et al. 1995;
     Ruiz--Lapuente, Canal \& Burkert 1995,1997; 
     Hachisu et al. 1995; Li \& van den Heuvel 1997; 
     Yungelson \& Livio 1998).

\bigskip

    The peak or characteristic time of those explosions is
    different: merging of WDs happens typically a few times
    10$^8$ yrs after star formation whereas the accretion of
    material from a non-WD companion involves less massive
    companions which take longer to leave the main sequence
    to become subgiants and giants. This second evolutionary
    path to SNe Ia takes about a few times
    10$^9$ yrs. Therefore, there should be a reservoir for
    SNe Ia in those stars 
    that evolve through a few Gyrs in cosmic
    time until they explode. 
    The overall expected number of SNe Ia as related to 
    star formation gives also an idea on the proportion of
    binaries that end up as Type Ia supernovae.
    
\bigskip

    Another factor entering in the evolution with z
    of the efficiency of Type Ia supernovae is the 
    role that metallicity plays in enhancing or 
    supressing the production of Type Ia supernovae
    from their parent stars.  If metallicity effects do not
    play any important role in SNe Ia,
    the number of SNe  Ia per unit mass in forming stars should
    remain insensitive to the progressive enhancement in metals
    of the interstellar medium along  cosmic history. 
    However, in the Algol--type scenario, it has been suggested that 
   metallicity plays an important role as the material transfered
   onto the WD forms a strong wind close to the 
   WD surface which prevents the accretion rates from becoming very large 
  (Hachisu, Kato \& Nomoto 1996). The role of this wind is claimed to be 
  very sensitive to metallicity, and calculations predict that  
  the number of SNeIa at low metallicity (z $>$ 1) should drastically
  decrease (Hachisu, Kato \& Nomoto 1996). 
   A drop at z $>$ 1 would thus confirm the metallicity effects
  suggested by Hachisu et al. (1998) and Kobayashi et al. (1998). 

\bigskip

    An important question is whether the efficiency in producing 
   SNe Ia has changed along cosmic time. 
     Results from several low--z  Type Ia supernova searches and 
     low--z star formation rate measurements do allow to address this issue. 
     Gallego et al. (1995) estimated the local star formation rate 
     from H$\alpha$ emission galaxies. The value derived from their
     search is $\dot \rho_{*} = 3.7\  10^{-2}$ $M_{\sun}$ h$^{2}$ Mpc$^{-3}$.  
     More recently, several authors have found values at least twice
     as large than the previous estimate for the local star formation
     rate (Gronwall 1999; Tresse \& Maddox 1998; Serjeant et
     al. 1998).
     Treyer et al. (1998) 
     from a UV-selected galaxy redshift survey find
     a local dust--corrected star formation rate of 
     4.3\ $10^{-2} $ M$_{\sun}$ h$^{2}$ Mpc$^{-3}$. If we take the
     rate obtained by  Hamuy \& Pinto (1999) from SNe Ia
     in the Calan/Tololo survey (SNe Ia $\sim$  2.2 10$^{-5}$ SNe Ia 
      yr$^{-1}$ Mpc$^{-3}$), and divide it by the local star formation rate,
     we obtain a local efficiency of SNe Ia out of star forming mass of 
    about 1.09 10$^{-3}$ M$_{\sun}^{-1} \ h_{65}^{2}$. 
     The efficiency seems to be similar up to z $\sim$ 0.55, if we take
     the rates by Pain et al. (1999) (see Table 1).  Possibly such
     high efficiency extends up to z $\sim$ 1, since 
      Aldering, Knop and Nugent (1999) anounce a similar rate 
     of SNeIa per comoving volume at z $\sim$ 1.

     An efficiency of SNe Ia out of star forming mass of
     $\langle {\cal E}_{SNeIa} \rangle = 1.41 \pm 0.35 \ 10^{-3} \
     M_{\sun}^{-1} \ h_{65}^{2}$, as derived taking into account
     rates measurements at various z  (see Ruiz--Lapuente
     \& Canal 2000), means that less than 
     1000 M$_{\sun}$ going into star formation
     give 1 Type Ia supernova (720 $\pm$ 250  M$_{\sun}$ give 1 SNIa).
     There is a delay between star formation
     and supernova explosion that makes this estimate move up or down
     according to the timescale and star formation history. But 
     most star formation histories would average out the effect since the 
     time distribution of SNe Ia is broad.
      Thus, the above number derived from observations 
     is physically meaningful.    
     We can work out the efficiency of Type II SNe explosions and
     compare it with the previous result. Assuming that all stars with
     M $\gapprox$ 10 M$_{\sun}$ do produce gravitational collapse 
     supernovae, and the Salpeter IMF, there is one gravitational 
     collapse supernova per 135 M$_{\sun}$ in forming stars, and
     thus the proportion of SNe Ia to SN II/Ibc supernovae is 
     roughly one to five.

    The above ratio of Type Ia over Type II SNe 
    is basically in agreement with
    chemical evolution arguments (Timmes et al. 1995).
    At z $\sim$ 0 one expects
    an increase of the efficiency for the two main candidates
    to SNe Ia as a result of their explosion timescale
    (significant formation of SNe Ia progenitor 
    systems occurred at high--z 
    and they only exploded now). The trend 3.5 Type II/Ib to 1 Type Ia
    in the local sample of spiral galaxies (Cappellaro et al. 1997)
    would be in agreement with those expectations. 
    As it is argued in Ruiz--Lapuente $\&$ Canal (2000), 
    the high efficiency in producing Type Ia supernova explosions
    out of star formation favors the single--degenerate scenario
    for Type Ia supernovae.

\section{ Gamma--ray emission} 

    Before integrating along cosmic time all the $\gamma$--ray 
   contribution from SNe Ia, we must examine the evolution
   of the emission with time as the supernova ejecta expand and
   become optically thin. Whereas the optical luminosity of
   supernovae peaks within three weeks after explosion, the 
   $\gamma$--ray emission achieves a maximum around two and a half months
   after explosion for Type Ia SNe and several months later for 
   Type II/Ibc SNe. The exact moment of the peak depends on the kinetic 
   energy, total mass of the supernova envelope and distribution
   of the radioactive material. As the Compton scattering optical
   depth of the  ejecta decreases due to the thining out of the material, 
    the $\gamma$--ray spectra of Type Ia and 
   Type II/Ibc SNe 
    start to show more prominent lines and less continuum emission. 
    The lines reveal the fractions of radioactive isotopes
     along with their mean half life.

  The radioactivity of SNe Ia is provided mostly by the decay 
   $^{56}Ni  \rightarrow   ^{56} Co \rightarrow ^{56}Fe$, which gives
   rise to $\gamma$--ray photons in several lines and positron 
   emission. The strongest lines 
   of $\rm ^{56} Co$ decay are at 0.847 and 1.238 MeV followed in strength 
   by the lines at 
   2.599 and 1.771 MeV and others (Browne et al. 1978). 
   The $\rm  ^{56}Ni$ decay has its strongest lines at 158 and 812 keV. 
   As this isotope, $\rm ^{56}$Ni, has a shorter mean half life
   (6.1 days) than its product decay $\rm ^{56}$Co (78.8 days),
   the latter is responsible for the SNe radioactivity
   through most of the SN life, i.e., its Comptonized lines give rise to the
   continuum emission for a few years.   
   $\rm ^{57}$Co (a less abundant 
   product of the explosive nucleosynthesis) has a mean half life of 271 days,
   and its emission is relevant for Type II SNe as it powers
   the bolometric light curve of those SNe for years.

   When     
    surveying the gamma--ray sky and integrating to obtain the 
    diffuse emission, we detect the emission from supernovae in 
    various phases, including those corresponding to  the time
    when supernovae have  already faded away in the optical.
    At a given moment, we see in the 
    $\gamma$--ray domain, from every  
    volume of the sky, the SNe Ia that were exploding at the
    epoch corresponding to the redshift of that volume, and also those 
    that exploded in that same volume up to almost two years before.
    We add up the emission of the supernovae at different phases.

   The calculation of
   the $\gamma$--ray spectra in this work 
   is done through Monte Carlo computations following the
   $\gamma$-ray emission arising in different regions of the 
   supernova envelope. The numerical algorithm includes as main 
   matter--radiation interaction processes
   Compton scattering, pair production and photoelectric effect. 
   The formation of positronium and its two and three 
   photon decay is included as well. Monte Carlo calculations
   of a similar kind were done for
   fast deflagration models
   (Ruiz--Lapuente et al. 1993) and for the 
   deflagration model W7 by Nomoto, Thielemann \& Yokoi (1984).
   Results on this standard Type Ia supernova model
   were compared with calculations by 
   Burrows \& The (1990) and Shigeyama et al. (1993), giving
   agreement among the predictions.

    SNe Ia, since they synthesize most of the $^{56}$Ni in the
    universe, are the most important contributors to
    the $\gamma$--ray background. SN II and SN Ibc contribute in a 
    lesser extent because of their lower proportion of radioactive 
    nuclei of medium range lifetime and their larger ejected mass and smaller 
    expansion velocities. Type II SNe and Type Ibc SNe produce a
    much more Comptonized $\gamma$--ray spectrum. They give rise to emission
    shifted to the hard X--ray domain.

     It is already known that Type Ia supernovae show a degree 
    of diversity in the amount of $^{56}$Ni synthesized in the 
    explosion (Ruiz--Lapuente \& Filippenko 1992; Ruiz--Lapuente \&
    Lucy 1992). 
    The spread in brightness at maximum of those SNe Ia is linked
    to the $^{56}$Ni mass as well as to 
    the effect of the opacity of the supernovae envelope
    (Arnett 1999; Pinto \& Eastman 2000). The distribution of
    Type Ia supernovae is centered, however, in ``normal'' events with a
    frequency of about 83--67 $\%$ according to local studies
    (Branch, Fisher \& Nugent 1993; Li et al. 2000). 
    Those are well reproduced
    by the explosion of a C+O WD which synthesizes
    about 0.6 M$_{\sun}$ of $^{56}$Ni. The deflagration explosion model W7 
    by Nomoto, Thielemann \& Yokoi (1984), that yields that amount 
    of radioactive Ni, has the characteristics
    in velocity structure, and element distribution that reproduces 
    well the prototype SNeIa. Therefore, we use that model to compute 
    the $\gamma$-ray emission. 
    Recently, some support for deflagration explosion models 
    has come from 3--D hydrodynamic calculations
    of SNe Ia explosions which favor such burning mode against  
    detonations and pulsated
    delayed detonations (Reinecke, Hillebrandt \& Niemeyer 2000).      
    The $\gamma$--ray emission of the modeled SNIa (model
    W7) resulting of an integration 
    of the spectrum over 600 days (a longer integration would not 
    add up significantly to the total emitted flux) can
    be seen in Figure 3 (see as well Figure 4). 

    The result of integrating over 600 days 
    the spectrum of a Type II/Ibc collapse supernova  can be seen for
     comparison in Figure 4. 
    The model used for the Type II supernova is a
    Type II core collapse that produced  
    0.175 M$_{\sun}$ of $^{56}$Ni and ejected an envelope of 
    14  M$_{\sun}$ (Eastman 1998). 
      Figure 5 shows the relative contribution of SNe II and SNe Ia. 
    In all the figures on $\gamma$--ray background calculations compared
    with observations, we have 
    included the diffuse $\gamma$--ray background produced by both
    SN types. We found, however, 
   the SN II contribution to be negligible 
    in the MeV range as compared with that provided by SNe Ia explosions. 
   The corresponding Type II SNe rates as compared to Type Ia SNe
    rates are shown in Figure 6.

    We have computed as well 
    the integrated emission for a Type Ia supernova coming from 
    a faster deflagration of a C+O WD (Ruiz--Lapuente et al. 1993).
    The explosion synthesizes 
    a larger amount of $^{56}$Ni. The flux level would be higher
    than for model W7 at a 10 $\%$ level. The early emission would
    reveal stronger $^{56}$Ni lines and a higher flux in the 
    keV range than model W7.
    This is an imprint of 
     SNe Ia explosions with  $^{56}$Ni extending close to the 
    surface (a mixed W7 or other model with 
    a larger spread of the radioactive material over the supernova 
    envelope would be stronger in the X--ray range as
    well). However, as 
    overluminous
    SNe Ia are in the overall sample
    compensated by the existence of the population of
     underluminous SNe Ia with a similar proportion
    in rates (Li et al. 2000), and those subluminous SNe Ia give 
    a lower $\gamma$--ray flux, the simplification of using just a model for 
    ``normal'' Type Ia SNe, on the overall, seems justified.

\bigskip

\section{Integration}

\bigskip

In order to calculate the contribution of SNeIa to the $\gamma$--ray background
spectrum in the MeV range, we first compute standard SNeIa $\gamma$--ray 
emission spectra for different times, from explosion to almost complete 
extinction of the produced radioactivities (about 2 years after explosion). 
As explained in the preceding paragraph, we compute the spectra 
from the widely used  W7 model (Nomoto, Thielemann, \& Yokoi 1984) (see 
again Figure 3, where the fluxes correspond to a distance of 10 Mpc). A 
time--averaged spectrum is used as the standard source for the 
background calculation.

If we call $f(E)$ the source flux at energy $E$, the background flux $F(E)$
from the whole sky is obtained by addition of the contributions from the 
SNeIa at different redshifts $z$:

$$F(E) = \int_{0}^{z_{lim}} SN\!R(z)\ f\left[E(1+z)\right]\ d_{L}(z)^{-2}\ 
(1+z)\ dz\eqno(2)$$

\noindent
where $SN\!R(z)$ is the SNeIa rate (corrected for time--dilation) in
the volume element between $z$ and $z+dz$ (calculated as in Ruiz--Lapuente \& 
Canal 1998, for the different SNeIa scenarios and SFRs), $d_{L}(z)$ is the 
luminosity distance corresponding to each of the three model Universes 
considered here, and the $(z+1)$ term accounts for compression of the energy 
bins. We have adopted a value $z_{lim}$ = 2.5 after checking that the 
contributions from SNeIa at higher redshifts is negligible.
Figure 7 shows the integrated  contribution up to various z from 
SNe Ia $\gamma$--rays. 
 Of course, $F(E)$ 
from (2) has to be multiplied by a constant adjusted to the distance for 
which the source spectrum has been calculated. The luminosity distance 
$d_{L}$ is given by:
 
$$d_{L} = { (1 + z) \over H_{0}|\Omega_{k}|^{1/2}}\
sin\!n\left\{|\Omega_{k}|^{1/2}\int_{0}^{z_{1}}\left[(1+z)^{2} (1+\Omega_{M}z)
- z(2+z)\Omega_{\Lambda}\right]^{-1/2}dz\right\} \eqno(3)$$

\noindent
where $\Omega_{k} = 1 - \Omega_{M} - \Omega_{\Lambda}$, and $sinn$ stands for 
$sinh$ if $\Omega_{k} > 0$ and for $sin$ if $\Omega_{k} < 0$ (both $sinn$ and 
the $\Omega_{k}$ terms disappear from (3) if $\Omega_{k} = 0$, leaving only 
the integral times $(1 + z)/H_{0}$, normalized to c=1) (Weinberg, 1972;
 Carroll, Press, \& Turner 1992). We will
see, however, that the resulting $\gamma$--ray background spectra are only
weakly sensitive to the model Universe adopted. Everywhere we take $H_{0} = 
65\ km\ s^{-1}\ M\!pc^{-1}$.

To the contribution of SNeIa to the $\gamma$--ray background spectrum, that 
from gravitational--collapse SNe (SNeII, SNeIb/c) has been added.
 When now computing the $SN\!R(z)$ used in
(2), for the different SFRs, we assumed that all stars with initial masses 
$M\geq 10\ M_{\odot}$ undergo gravitational collapse, with production of
 a SN II/Ibc. In the next section we present the results obtained for
 different SFRs.

\bigskip

\section{Results}

\bigskip

\subsection{Supernovae and Cosmic Gamma--ray Background}

The inspection by Watanabe et al. (1999b) of the
supernova contribution to the $\gamma$--ray background  
left open the question as to whether supernovae are indeed 
responsible for
this background. According to their work, 
  in the case of: i) a SNIa
timescale to explosion longer 
than 1 Gyr, and/or ii) the peak of star formation occurring at a
redshift much larger than 1, an alternative explanation to
 Type Ia supernovae as the source of the diffuse cosmic  
$\gamma$--ray background would be needed.  
However, SNe Ia rates used by those
authors are  below that observed at high-z. Moreover, the 
constrains derived on the timescale of the SNe Ia to reach
explosion after star formation do not hold when one looks 
to the derived efficiency for making SNe Ia out of 
star formation. 

 As they clearly point out, larger SFRs than the one used by them
 would take away the discrepancy and reproduce the observed
 $\gamma$--ray background. 
 Their model of SF history  sharply decreases  
 at z $ >$ 1 (they use a simplified SFR  
 first introduced by Yungelson \& Livio (1998) to model 
 Madau's early SFR rate) and it seems
 to underestimate  the high-z star formation rate.

  We find in the present work that there 
  is a range  of evolution of the SFR which would
  allow to account for both the  measured 
  Type Ia supernova rates and the
  observed cosmic $\gamma$--ray background. Figures 8--14
  show the results from calculations done with different star formation rates.
  Both rates of SNe Ia per comoving volume and their corresponding
  integrated $\gamma$--ray background are displayed. The SNe Ia rates
  are calculated for the coalescence of WDs, and for the Algol--type
  scenario including metallicity effects (Hachisu et al. 1996; 
  Ruiz--Lapuente \& Canal 1998). Due to the good agreement of the 
  efficiency of SNe Ia out of star formation on the later scenario, 
  this is the one used for the extragalactic $\gamma$--ray background. 
  Some SFR fall short
  both to explain the extragalactic $\gamma$-ray background 
  and the cosmic SNe Ia rate (Figures 8 and 9), and other clearly exceed the 
  level of SNe Ia contribution as compared with the optical rates
  and the $\gamma$--ray data. Among those overexceeding SFRs 
  we show a sharply peaked star formation rate compatible, 
  according to Blain et al. (1998), with all submillimeter
  and far--infrared data (gaussian peaked SFR referreed
  here to as Blain98b in Figure 1), having a maximum at z$\sim$ 2. 
  It clearly overproduces cosmic $\gamma$--ray background as 
  compared with the observations (Figure 10),
  and at the same time it predicts very high SNe Ia rates.
  On the other side, 
  the early Madau et al. (1996) SFR falls short (as already discussed)
  and gives too low an increase of SNe Ia between the local
  and the high-z values. The same occurs with a model of SFR peaking at
  z $\sim$ 2 compatible with the SCUBA results (Hughes et al. 1998), as
  can be seen in Figure 9. 
  Other models compatible with the 
  SCUBA results showing a SFR roughly constant after z $\sim$ 1
  (model Blain98 in Figure 1)
  give results compatible with the SMM and COMPTEL observations (Figure 11) 
  Their associated SNe Ia rates  nicely fall into the observed rates 
  of SNe Ia measured at z $\sim$ 0.4 and 0.55 (Pain et al. 1996;
  1999). The SFRs derived from the star
  formation history of Lyman break galaxies (Steidel et al. 1998)
  and the one from ISO (Dwek et al. 1998) go in the same sense
  for the predictions of rates and $\gamma$--ray background as the
  model by Blain et al. (1998). However, a slight increase of the
  absolute level of the SF would provide a better agreement with 
  the observations (Figures 12 and 13).  
  The SFR compatible with the EBL as detected through COBE
  also gives a reasonable fit to the data (Figure 14). 
 
  Two cosmologies have been used when integrating
  the $\gamma$--ray background: an open universe with $\Omega_{M}$=0.3, and
  a flat universe with $\Omega_{M}$=0.3 and 
  $\Omega_{\Lambda}$=0.7. Both the integrated 
  $\gamma$--ray background 
  and the SNe Ia counts at z $\sim$ 1 are larger in
  a $\Lambda$--dominated universe. The counts are, however, 
  much more sensitive 
   to the cosmological model (see Ruiz--Lapuente \& Canal 1998).

\subsection{Dust} 

    The role of dust obscuring Type Ia SN at high--z 
   can also be tested by comparing the extragalactic 
   $\gamma$--ray background expected from SNe Ia and the observed
   SNe Ia counts at high z. 
  
   If dust plays a very important role obscuring 
   Type Ia supernovae up to z $\sim$ 1 and leading
   us to miss their detection,  we should see a 
  discrepancy between the derived results from the
  optical searches and the level of the $\gamma$--ray 
  flux. The $\gamma$--ray background, which is not
  disturbed by dust absorption, clearly points out that 
  we can not have a much higher rate of SNe Ia
  than what we are observing optically. The reason is
  that with a much larger rate the $\gamma$--ray 
  background from those SNe Ia would exceed the fluxes
  measured by SMM and COMPTEL (see Figures 8--14).
  It might however be the case that future SN searches at 
  z$>$2 with NGST would reveal significant obscuration in the
  SN counts, more significant at those z  than below 
  1.5. It looks as if  the dust, being present 
  at those z, would redden the supernovae  by
  a  magnitude increment which does not affect significantly 
  the efficiency of the optical searches.

\bigskip
\bigskip

 In order to quantify this effect, we 
 have estimated the extinction by dust on the observed SNeIa 
counts. To that purpose we adopt the model of Pei, Fall, \& Hauser 
(1999) and take their redshift distributions of gas density 
$\Omega_{g}$ and of metallicity $Z(z)$. From that, we compute the dust 
distribution $\Omega_{d}(z)$ adopting a mean dust--to--metals ratio 
$d_{m} = 0.45$. The dust extinction and absorption 
coefficients, $\kappa_{e}(\lambda)$ and $\kappa_{a}(\lambda)$, as functions 
of wavelength $\lambda$, are those for LMC grains (Draine \& Lee 1984; Pei
1992), and 
the characteristic optical depth of absorbers $\tau_{t}(\nu, z)$, and
the characteristic optical depth of small--scale dust clumps 
$\tau_{c}(\nu, z)$, are computed accordingly. 
 Once those optical depths are obtained, we estimate the mean fraction of
photons absorbed by dust $A_{\nu}$,
calculate the extinction for the central wavelength of the 
$R$--band (5500 \AA) as a function of $z$, and translate it into a magnitude
increase $\Delta m_{R}(z)$.

To obtain an estimate of the effect on counts, we have computed the cumulative
SNeIa counts (SNeIa per yr and per sq.deg.) as a function of limiting magnitude
$m_{R}$ as in Ruiz--Lapuente \& Canal (1998). Comparison of the expected
counts with and without dust extinction is shown in Figure 15. The dashed line
shows the smaller counts resulting from the decrease in $z$ for a given 
$m_{R}$ when dust extinction is taken into account. The effect is rather small 
and only noticeable for $21.5\leq m_{R}\leq 23.2$. That reflects the double 
dependence of $\Omega_{d}$ on metallicity $Z$ and on gas fraction 
$\Omega_{g}$. At smaller $m_{R}$ (lower $z$), metallicity is high but there is 
little gas and thus $\Omega_{d}$ (proportional to $\Omega_{g}$) is small. At 
larger $m_{R}$ (higher $z$), there is more gas but then $Z$ is low, which also 
results in small $\Omega_{d}$. The overall effect in a magnitude limiting 
 SNe Ia search is that the fraction of SNe Ia missed due to dust obscuration
 in their host galaxies 
will not exceed a 5$\%$ in the redshift range (till z $\sim$ 1-1.5) 
 relevant for this work. This might explain the extraordinary consistency 
 between optical and $\gamma$--ray results.

\bigskip

\subsection{Other sources or supernovae?} 

   Unlike the other candidate sources proposed to explain
   the $\gamma$--ray background in the MeV range --i.e. 
   MeV blazars, Seyfert with MeV nonthermal tails-, 
   Type Ia supernovae have an evolution of rate
   with z that has been measured (Table I). 
   Knowing the SNe Ia rate at various z we only
   need to trust our  expectations for the spectra 
   in the $\gamma$--ray domain
  to be able to firmly establish our conclusions. 
   The $\gamma$--ray predictions from models do not raise
   major questions concerning the physics needed to be included 
   to calculate the spectra. That is basically the same
  physics needed to explain Type II SNe $\gamma$--ray emission,
   where the comparison with the $\gamma$--ray observations of SN 1987A
   proved successful (Sunyaev et al. 1988; Cass\'e \& Lehoucq 1994; 
   Nomoto et al. 1994). The SNIa explosion models have, in addition,
   been tested and compared with  
   optical spectra. While detection of individual SNe Ia 
   in the $\gamma$--ray domain has to await the launch of
   INTEGRAL and would only be possible in a certain extragalactic 
   distance range
   (Kumagai \& Nomoto 1997; Isern et al. 1999), here we
   outline the already available 
   detection of SNe Ia in the form of their observed MeV background. 
   Support for the identification of 
   the underlying radioactivity responsible for the optical light curves
   in SNe Ia comes from accurate tests.  
   The amount of radioactive material 
   synthesized in a SNe Ia is confirmed, among others,
   by the observational calibration 
   of the optical absolute luminosity of those explosions done 
   by means of Cepheids (Sandage et al. 1998; Gibson et al. 1999).

   Whereas it is worth exploring whether 
   Seyfert galaxies present a nonthermal tail in the
   MeV range as that observed in ``microquasar'' sources,
   and the blazar emission
   at those energies should be investigated, we find that the shape
   and flux of the $\gamma$--ray background is at the level
   expected from Type Ia SNe. The reconstruction of the distribution 
   with z of SNe Ia events provides the basis for this assertion.

\subsection{Cosmic chemical evolution and the SFR}

\bigskip 
\bigskip 

We have shown how the SNeIa fit to the cosmic $\gamma$--ray background
 in the MeV range probes the global
star formation history up $z\simeq 1.5$. It seemingly disfavors
 SFRs which rise 
only moderately from $z = 0$ to $z\simeq 1.5$ having a maximum star formation
value at that z below 0.2 M$_{\odot}$ yr$^{-1}$ Mpc$^{-3}$. That has 
important implications
for the overall cosmic chemical evolution, and particularly, 
for the history of metal 
enrichment as well as for destruction of fragile primordial nuclides, namely 
 deuterium. Cass\'e et al. (1998) discuss different galactic chemical 
evolution models and compare them with the data on the variation of the 
luminosity density, as a function of $z$, in different wavelength bands (UV, 
B, IR). They assume that the star formation history of our Galaxy is 
representative of the average galaxy that contributes to the luminosity 
density at high $z$, and they also find that, in order to fit the multi--color 
luminosity data for $0\leq z\leq 1$, SFRs rising sharply (by factors 
$\sim 10$) along this redshift interval are required. Spiral galaxies like 
our own should give a dominant contribution to the UV luminosity density up 
to $z\simeq 1.5$ (Marzke et al. 1994; Driver et al. 1999), and thus their 
star formation and chemical evolution histories should approximate well the 
global ones. As already pointed out above, Dwek et al. (1988) equally need 
high SFRs at high $z$ to reproduce the strong intergalactic IR background, 
and we have seen that their proposed $\rm SFR(z)$ gives a 
good fit to the cosmic $\gamma$--ray background.

Concerning the global evolution of metallicity with $z$, chemical evolution
models typically predict a stronger evolution than that observed. This is
the ``missing metals'' problem (Pagel 1998; see also Pettini 1999). Damped
Lyman alpha systems, which were thought to be the progenitors of today's
spirals, do not show significant chemical evolution, but the most
likely explanation is that Ly$\alpha$ systems trace a population different 
from the galaxies responsible for the bulk of star formation. At higher $z$, 
the metal contents of the Ly$\alpha$ forest, which may account for a large 
fraction of the baryons in the Universe, is one order of magnitude too low, 
but the missing metals can be in hot gas in galactic halos and 
proto--clusters. The role of the metal enrichment by SNe Ia at high z
is another unknown in those studies, and hardly constrained so far. 
The present combined results of the rate of SNe Ia at high z as 
observed in the optical and in the $\gamma$--ray background places 
constraints which are worth considering in models of cosmic chemical 
evolution. The role of SNe Ia generously enriching the Universe in metals
after z$\sim$ 1 can now be quantified and
compared with the observations. Such a high enrichment 
should show up in a steep change of the abundance slope 
of metals produced in SNe Ia  at z $\sim$ 1 (further discussion 
on this subject will be addressed in future work).

As to D, the abundances inferred from high--$z$ QSO absorbers show a 
strongly bimodal distribution, one value being ``low'' ($\rm D/H\simeq 
3\times 10^{-5}$: Tytler et al. 2000) and the other ``high'' 
($\rm D/H\sim 10^{-4}$: Webb et al. 1999). Taken as the
 primordial values, since 
D is only destroyed by astration in the course of galaxy evolution, the 
``low'' value (not much higher than the present one) would mean little 
destruction and thus scarce astration, corresponding to a stellar formation 
activity that would only moderately increase with lookback time, 
as considered often in  models for 
our Galaxy. In contrast, the  $\rm SFR(z)$ required to explain the cosmic 
$\gamma$--ray background as well as the observations mentioned 
 in section 2, would destroy primordial
D by a large factor ($\sim 10$). Overall consistency would thus require a 
``high'' primordial D (which means a low baryon to photon ratio: 
$\eta_{10} = 2-3$ in the standard BBN, higher values being possible
in non--standard BBN), followed by its massive destruction by astration down
to the present low values (Vangioni--Flam \& Cass\'e 1995).

\section{Conclusions}

 Both the spectral shape and the level of the 
 overall $\gamma$--ray flux
 escaping from Type Ia supernova explosions up to z$\sim$ 1
 match  the measurements of the  
 cosmic $\gamma$-ray background obtained with
  SMM and COMPTEL in the window
 covering the hard--X ray range till 3 MeV
 (Kappadath et al. 1996;
 Watanabe et al. 1999b).
 
 The results have been checked against the empirical 
 high-z SNe Ia rates obtained in searches designed for cosmological purposes.
 The analysis of the supernova rates both in the high--z and 
 in the nearby universe allows us to depict the 
 efficiency in producing SNe Ia. Once the SNIa
 efficiency from star forming mass is known at various z, 
 both the SNe Ia rates per comoving volume and the absolute level 
 of the $\gamma$--ray background constrain the SFR at high z. 
 In this work we have found that a range of SFR histories is compatible
 with these observations. The best results are obtained for 
 SFR reaching a maximum value of 0.3--0.4 M$_{\sun}$ yr$^{-1}$ Mpc $^{-3}$
 at z $\sim$ 1.5, and possibly keeping that value towards higher z
 (see Figure 10). Some of the reconstructed star formation histories
 compatible with the SCUBA measurements (Blain et al. 1998) are
 among those favored by this analysis. 
 The SFR derived from the UV emission density at high z 
 in the Hubble Deep Field by Madau, Pozzeti \& Dickinson (1998) seems 
 a factor 2 too low to explain the SNe Ia counts and the
 COMPTEL and SMM background observations in the MeV range.  Other SFRs
 peaking with a maximum of star formation 3-4 times larger than 
 the above value can be excluded on account of the
 overexceeding $\gamma$--ray background. The SFR derived from COBE
 (Dwek et al. 1998) gives results in background and
 rates compatible with the observations.

  Beyond considerations on the SFR it is shown in this work that
  for a rate evolution of SNe Ia compatible with a
  number of empirical determinations in searches performed by
  the Supernova Cosmology Project (z $\ge$ 0.3), the EROS SN search 
  (z $\sim$ 0.1), and others (Table 1), we obtain a $\gamma$--ray background
  that explains the spectral shape and flux level measured
  by COMPTEL and SMM  (Kappadath et al. 1996; Watanabe et al. 1999b; 
  Weidenspointer 2000).   
  Earlier suggestions of a too low contribution by supernovae
  in the MeV range (Watanabe et al. 1999b) implied as well constrains
  on the delay time between formation and explosion of SNe Ia
  that are not substained in the present analysis. We found that while
  it is still early to constrain the timescale to explosion of SNe Ia,  
  the efficiencies
  in producing SNe Ia supernovae out of star formation, 
  as discussed here, provide a reliable basis for a $\gamma$--ray 
  background calculation (see Ruiz-Lapuente \& Canal 2000 for the physical
  constrains on the progenitors). It is clearly seen that
  a range of star formation histories compatible with current
  measurements (SCUBA,COBE) reproduce well both the SNe Ia rates
  and the $\gamma$--ray background.

  In view of the presence of 
 sources other than supernovae contributing to the background 
 in the keV region and above 3 MeV, a finer spectral study of
 this cosmic $\gamma$--ray emision appears compulsory.  
 There are several interesting goals for 
 future $\gamma$--ray missions in the MeV range. 
 Higher spectral resolution and more precise  
 flux determination 
 in the measurement of the extragalactic MeV 
 background will be most important, since they will allow to check
 the consistency among cosmological model, history of star formation
 and Type Ia supernova efficiency which are being derived 
 from the cosmological SNe Ia searches. 
 As already noticed (Watanabe et al. 1999b), the individual $\gamma$--ray
 lines when integrating the supernova emission along z add up to an
 observable  pattern of step structure in the background
 spectrum. Such pattern, unobserved in measurements up to the present 
 achieved accuracy, should be detected when performing high precision
 measurements of the cosmic 
 $\gamma$--ray background.
  Another question which can not be answered with the data available
 at this point is the 
  shape in the junction of the spectrum between the region where SNe Ia 
  provide the emission and blazars remain the main emitters. Next 
  missions with coverage of the 3-10 MeV region could be critical
  in providing the accurate measurements assesing the blazar contribution
  in this transition range.

  Back to  longer wavelengths, 
  future SN searches using the NGST 
  should provide  measurements
  of rates of SNe well beyond the critical peak at z$\sim$ 1--1.5
 (Madau 1998; 
  Dahlen \& Fransson 1998). 
  Those would not only be used to test the cosmological model in the way
  done up to now through optical SN searches 
  by the Supernova Cosmology Project and the High--z Team Collaboration
  (Perlmutter et al. 1999; Schmidt et al. 1998), 
  but they will provide as well  
  additional tests on $\Omega_{\Lambda}$ through the counts evolution
  with z (Ruiz--Lapuente \& Canal 1998), and will be able to answer
  a large number of questions related to the star formation history
  in the Universe, its chemical evolution and the background light 
  provided by its stellar explosions.

\bigskip
  
  PRL thanks her colleagues of the Supernova Cosmology Project, 
  and specially R. Pain, for exchanges on the observed high--z rates.
  The authors would like to thank as well R. Eastman for providing 
  his core--collapse models.

\vfill\eject

\vfill\eject

\section{Tables}

\begin{table*}[htb]
\caption{Type Ia supernova rates along z}
\label{table:1}
\newcommand{\m}{\hphantom{$-$}}
\newcommand{\cc}[1]{\multicolumn{1}{c}{#1}}
\renewcommand{\arraystretch}{1.2} % enlarge line spacing
\begin{tabular}{@{}lllll}
\hline
Redshift  & $\tau_{SNu}$ & $\rho_{Ia}$  & 
 counts & Search \\

$  < z>$   & SNu  h$_{0.65}$$^{2}$ & Ia Mpc$^{-3}$yr$^{-1}$h$_{0.65}$$^{3}$  & 
 Ia yr$^{-1}$$\Delta z ^{-1}$sqdeg $^{-1}$  & Search/Author \\
\hline
 0.     & 0.21$^{+0.30}_{-0.15}$ & 2.2$^{+3.4}_{-1.4}$ 10$^{-5}$ &  - 
   & Calan/Tololo$^{1}$  \\
        & 0.15$\pm$0.05   &  -    &  -  & 5 combined searches$^{2}$  \\ 
 0.1   & 0.12$^{+0.13}_{-0.08}$   &  1.7$^{+1.9}_{-1.1}$ 10$^{-5}$  &
     -         & EROS2$^{3}$       \\
 0.32   &  $<$ 0.32 (1$\sigma$)  & $<$ 4.52 (1$\sigma$) 11.02
 (2$\sigma$) 10$^{-5}$  
 &     & INT search $^{4}$    \\
     &                  & $<$ 6.2 (1$\sigma$) 15. (2 $\sigma$)
  10$^{-5}$ $^{*}$  & 
                      &    \\
 0.38    & 0.35$^{+0.38}_{-0.26}$ & 4.8$^{+3.3}_{-2.2}$ 10$^{-5}$  & 
      160.7$^{+111.7}_{-75.7}$ &  SCP $^{5}$  \\
     &                  & 6.9$^{+4.8}_{-3.2}$ 10$^{-5}$ $^{*}$  & 
                      &    \\
 0.55   & 0.25$\pm$ 0.08 & 4.53 $^{+ 1.43}$$_{-1.35}$ 10$^{-5}$  &
  81.0$^{+23}_{-21.8}$  & SCP $^{6}$ \\
        &    &6.74$^{+2.13}$$_{-2.00}$  10$^{-5}$ $^{*} $ &
      &        \\
\hline
\end{tabular}\\[2pt]
The rates for comoving volume are given for the cosmology $\Omega_{M}$
 =0.3 $\Omega_{\Lambda}$=0.7. Those rates marked with the asterisk
* are for $\Omega_{M}$ =0.3 $\Omega_{\Lambda}$=0. 
$^{1}$Hamuy \& Pinto (1999); $^{2}$Cappellaro et al. (1997); 
$^{3}$Hardin et al. (1999);$^{4}$Hamilton (1999);$^{5}$Pain et al.(1996);
$^{6}$Pain et al.(1999); see also Fabbro (2000). 
\end{table*}

\bigskip

\vfill\eject

\section{Figure captions}

\bigskip

\figcaption{Explored global $SFR(z)$, from Madau et al. (1998);
 Hughes et al. (1998); Blain et al. (1998) models compatible with
 the SCUBA results; Steidel 
 et al. 1998; Dwek et al. (1998) models compatible with COBE and ISO
 results.}

\bigskip

\figcaption{
 Predictions for the ``efficiency'' in producing SNe Ia
 at a given z per unit of mass in forming stars. The
 curves show the expected evolution of that efficiency 
 for two SNe Ia candidates systems with different 
 timescales to explosion: merging of double 
 degenerate pairs (plotted DD with a dashed line) and 
 and Algol--type binary pairs with wind effects included
 (plotted CLSW with a solid line). The data points have
  been derived using the SNe Ia measurements 
  (Pain et al. 1996, 1999; Hamuy \& Pinto 2000; Hamilton 1999;
 Hardin et al. 2000) and the star formation rates
  (Madau et al. 1998; Steidel et al. 1998; Blain et al. 1998).
   For the SFR used 
   details are given in section 6. Implications for the SNe Ia
  progenitors are
  presented in Ruiz--Lapuente \& Canal 2000. The quantity 
  plotted  allows a good identification of the SNIa progenitor since
  it is mainly sensitive to the efficiency of the  progenitor
  systems in giving Type Ia explosions. It only shows a negligible 
  dependence with the cosmological model and weak dependence on 
  details of the  history of star formation.}

\bigskip

\figcaption{
Top panel: Calculated $\gamma$--ray spectrum of
a Type Ia supernova integrated over 600 days.  
The model chosen is the deflagration of a 
C+O WD of 1.38 M$_{\sun}$ (model W7 by Nomoto, 
Thielemann \& Yokoi 1984).
Bottom panel: Calculated $\gamma$--ray spectrum of 
a Type II supernova integrated over 600 days.
 As compared with a Type Ia supernova, we see that 
the emission is mainly coming out below 100 keV
as a result of the strong Comptonization of the
$\gamma$--rays in the massive envelope of 
those exploding stars.}

\bigskip

\figcaption{Gamma--ray spectra from the previous models given
 per unit nucleus $^{56}$Ni.}

\bigskip

\figcaption{
Relative contributions of Type Ia and Type II SNe
to the $\gamma$--ray background. The Type II SNe 
contribution is negligible compared with the Type Ia SNe
 contribution in the MeV range.
The diffuse $\gamma$--ray background data are
from SMM (dashed line) from the analysis by Watanabe
 et al. (1999a), the filled square points are COMPTEL data as analysed
 by Kappadath et al. (1996), and the pentagons come from the COMPTEL
 analysis by Weidenspointer (2000). The starred symbols come from the
 reanalisis of the Apollo data (Trombka et al. 1977). }

\bigskip

\figcaption{
Rates of Type II and Type Ia SNe along z
for different star formation histories. While Type II SNe are more
frequent than Type Ia SNe, their contribution to the $\gamma$--ray
background is lower (see figure 6). Type II SNe trace the star
formation rate. Type Ia SNe at high z give an indication of the SFR,
the characteristic timescale since star formation for the evolution
of the progenitors till explosion, and the efficiency of those
to reach the explosion stage (Ruiz-Lapuente et al. 1995).   
}

\bigskip

\figcaption{
The $\gamma$--ray background  from SNe Ia adding up the emission
up to different z from those explosions. The star formation rate
by Madau et al.(1998) without dust correction
 is used as reference here, while it is
found (see next figure) to be a too low SFR.
}

\bigskip

\figcaption{{\it Top panel}: 
Prediction for the diffuse $\gamma$--ray
background from Type Ia supernova
 binary progenitor of Algol--type with
the efficiency shown in figure 1 (named CLSW), and the star
formation rate from Madau et al. (1998) without
dust--correction for the SFR beyond z $\sim$ 2.
The diffuse $\gamma$--ray background data are
from SMM (dashed line) from the analysis by Watanabe
 et al. (1999a), the filled square points are COMPTEL data as analysed
 by Kappadath et al. (1996), and the pentagons come from the COMPTEL
 analysis by Weidenspointer (2000). 
{\it Bottom panel}: Rates of Type Ia supernovae 
along z from the progenitor used in the above calculations
compared with the results of the high--z searches (Pain et al. 1996; 
Pain et al. 1999.). See all the rates listed in Table 1.  
The two lines in both panels show the results for 
the two cosmologies chosen: a flat Universe with $\Omega_{M}$=0.3
and $\Omega_{\Lambda}$=0.7 (dashed line) and an open 
Universe with $\Omega_{M}$=0.3 and  $\Omega_{\Lambda}$=0 
(solid line).}

\bigskip

\figcaption{{\it Top panel and bottom panel}:
Predictions compared with observational data
as in Figure 8 for the cosmic $\gamma$--ray 
background due to Type Ia supernovae and the corresponding
high--z rates. Here a broad SFR by Hughes et al. (1998) 
peaking at a z$\sim$ 2 and compatible with the SCUBA results 
is used.}

\bigskip

\figcaption{{\it Top panel and bottom panel}:
Comparisons of predictions from models for both the
cosmic $\gamma$--ray background (upper panel)
 and  Type Ia supernova counts (lower panel). 
The SFR used here corresponds to a SFR with a steep rise 
 to a maximum at z $\sim$ 2 (Blain et al. 1998, model Blain98b in
 Figure 1),
and it shows how that SFR is beyond the upper limit of 
what $\gamma$--ray observations allow and is incompatible 
as well with the results from SNe Ia cosmological searches.}

\bigskip

\figcaption{{\it Top panel and bottom panel}:
Comparisons of predictions from models for both the
cosmic $\gamma$--ray background (upper panel)
 and  Type Ia supernova counts (lower panel). 
The SFR used here corresponds to a SFR by 
Blain et al. (1998), model Blain98 in
 Figure 1. It is a SFR that stays constant for z higher than 
 2.}

\bigskip

\figcaption{{\it Top panel and bottom panel}:
Predictions and observations as in Figures 8-11 are
shown here for the SFR from Steidel et al. (1998).}

\bigskip

\figcaption{{\it Top panel and bottom panel}:
Predictions and observations as in Figures 8-12 are
shown here for the SFR from Dwek et al. (1998), compatible with 
 the results from ISO.}

\bigskip

\figcaption{{\it Top panel and bottom panel}:
Predictions and observations as in Figures 8-13 are
shown here for the SFR from Dwek et al. (1998), compatible with 
 the results from COBE.}  

\bigskip

\figcaption{Number counts of Type Ia supernovae in limited 
 magnitude searches at high z. Results for cumulative number 
 counts without taking into account obscuration by dust (solid line)
 and results taking into account the effect of extinction by dust (dashed 
 line). The dust model used is as in Pei, Fall \& Hauser (1999), and
 depends on the metal enrichment of the gas along z.}

\vfill\eject

\clearpage

\begin{figure}[hbtp]
\centerline{\epsfysize15cm\epsfbox{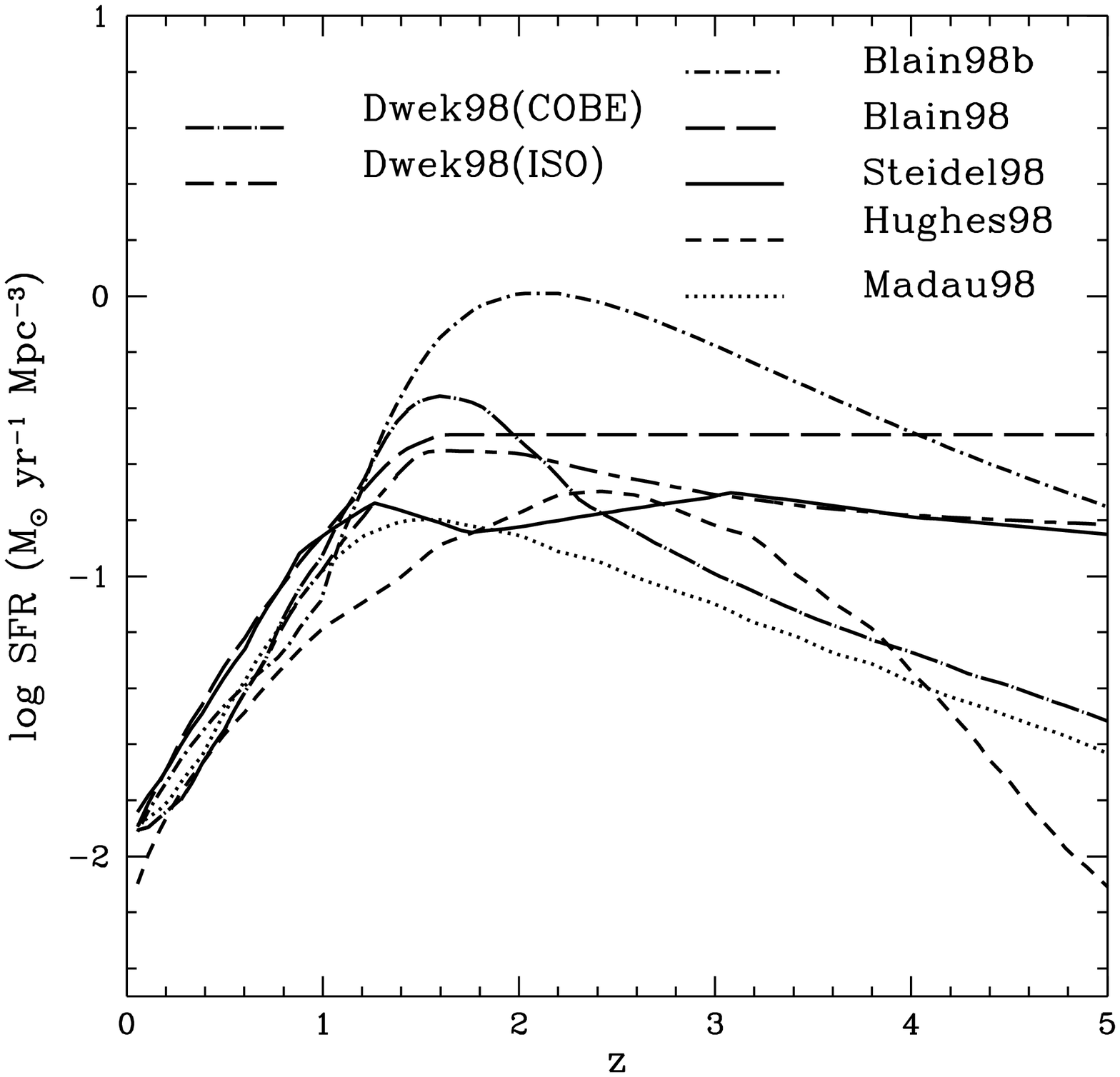}}
\nopagebreak[4]
\label{fig1}
\end{figure}

\clearpage

\begin{figure}[hbtp]
\centerline{\epsfysize20cm\epsfbox{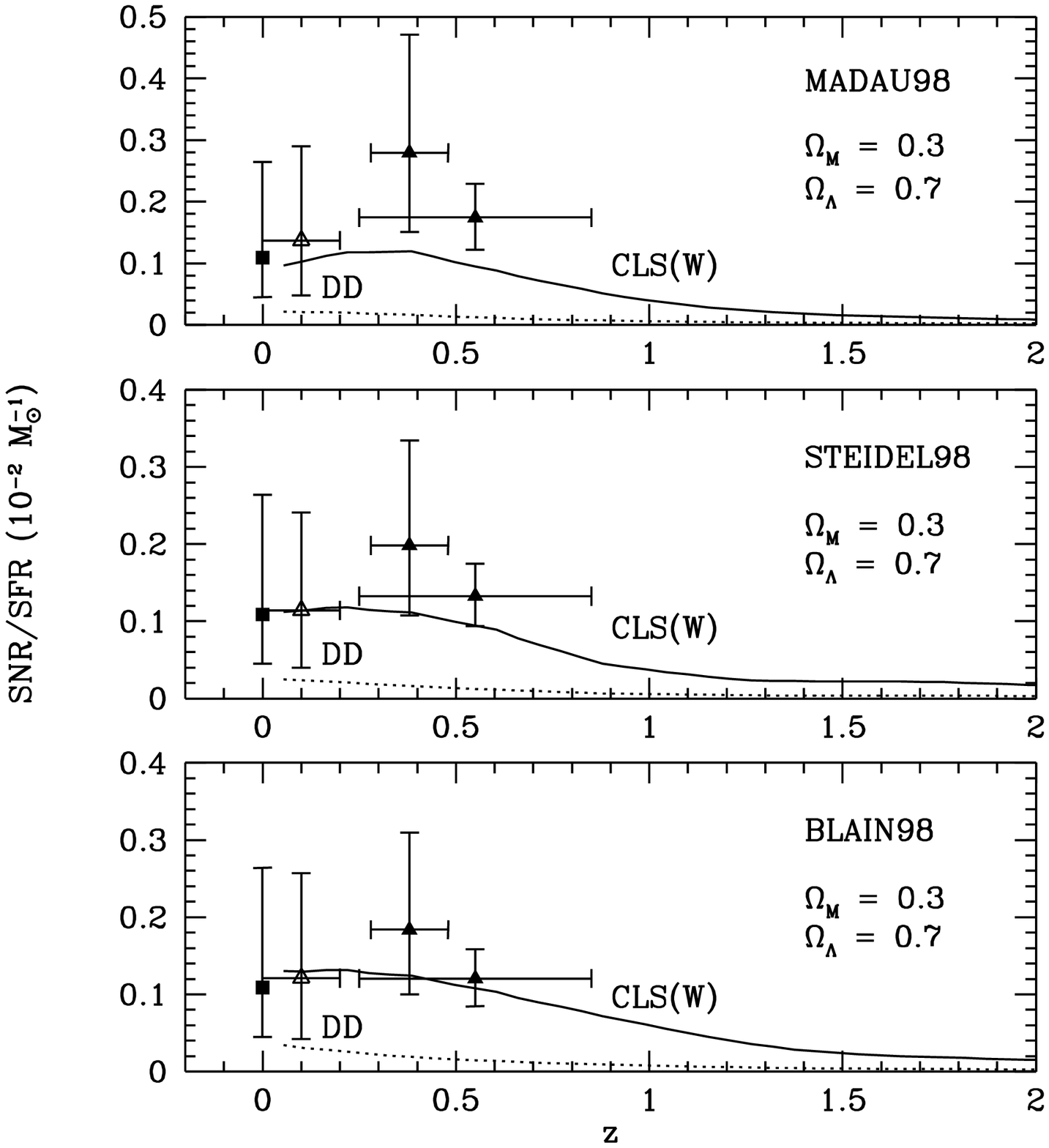}}
\nopagebreak[4]

\label{fig2}
\end{figure}

\clearpage

\begin{figure}[hbtp]
\centerline{\epsfysize10cm\epsfbox{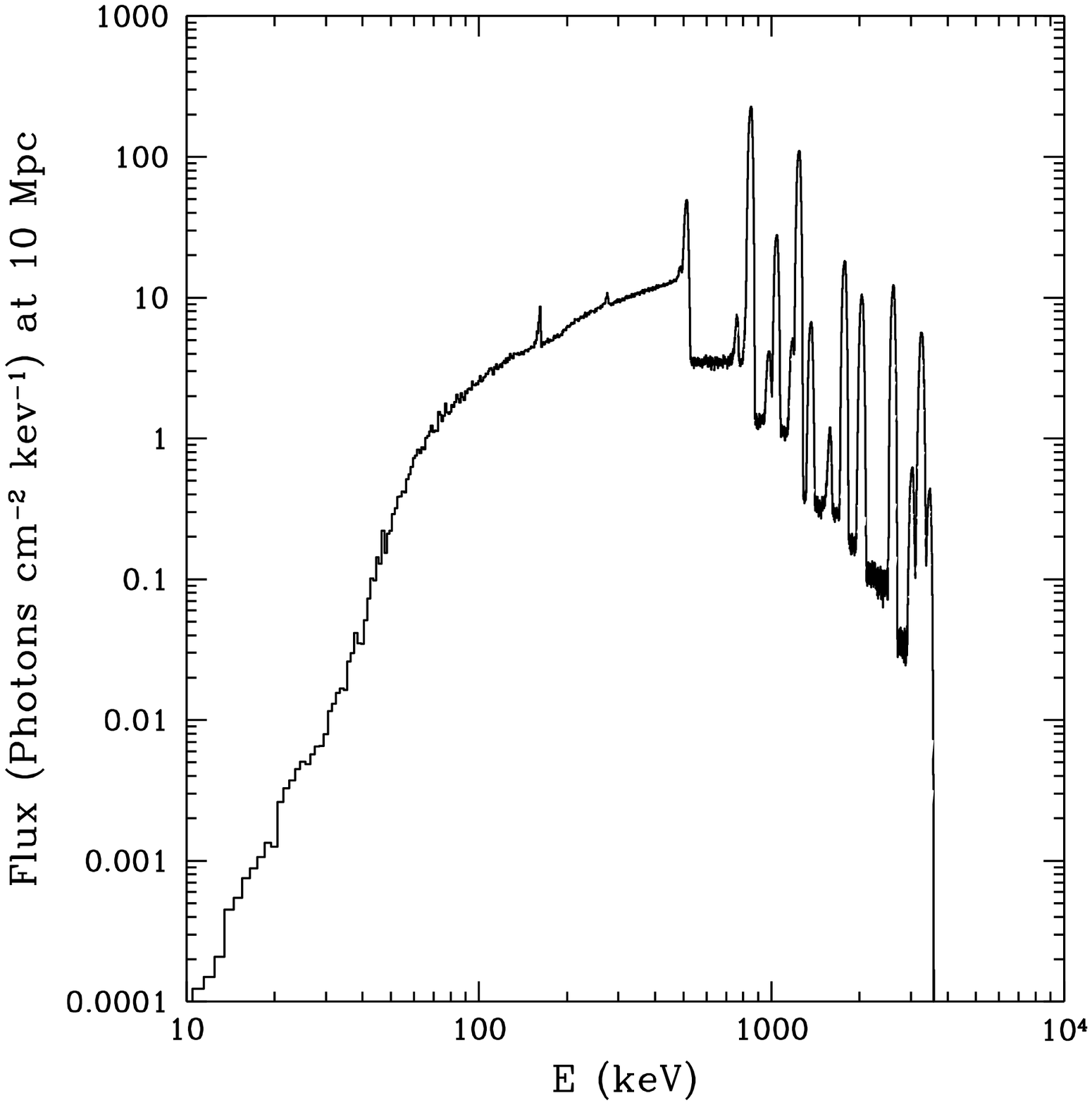}}
\nopagebreak[4]
\centerline{\epsfysize10cm\epsfbox{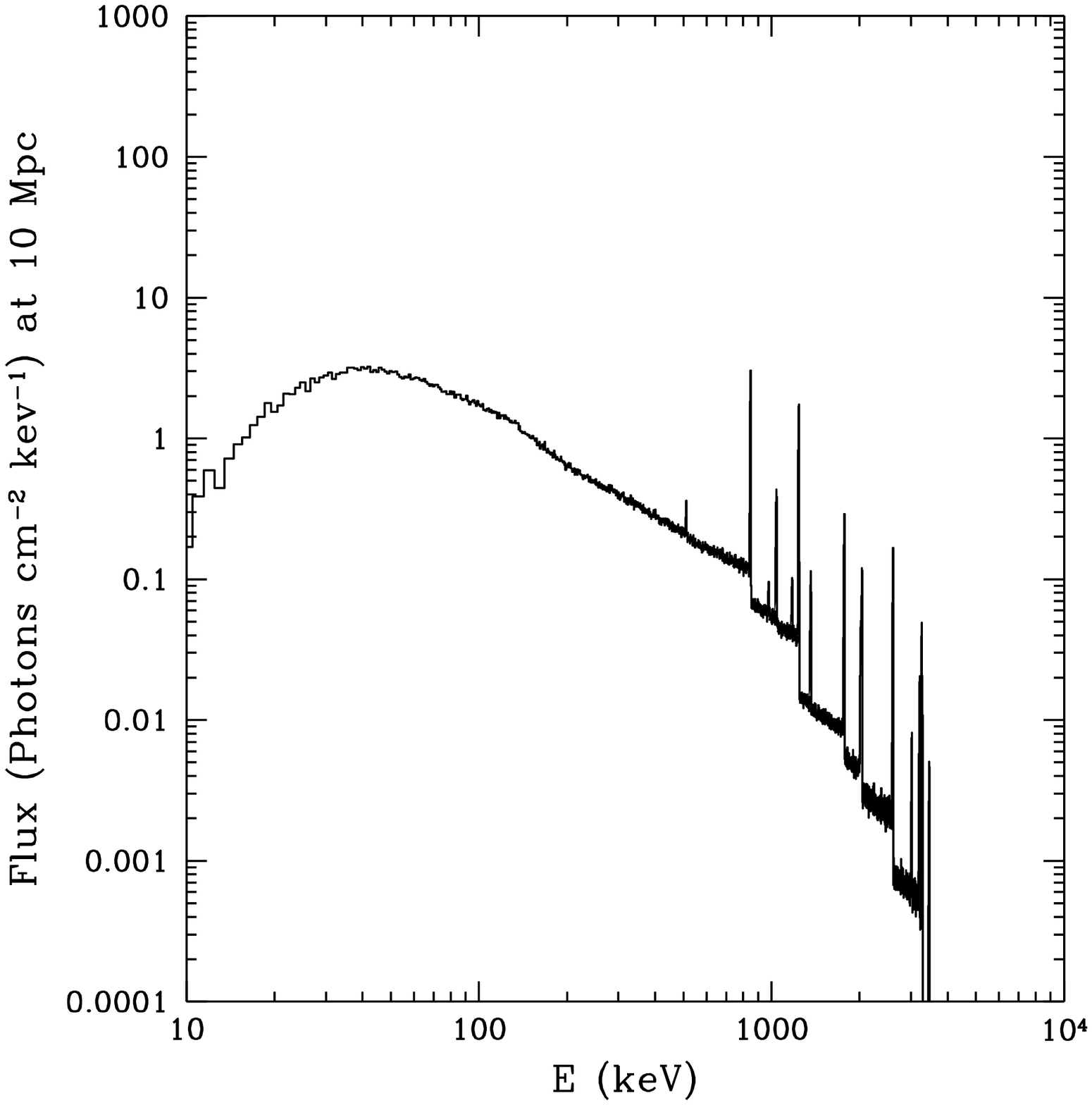}}
\nopagebreak[4]
\label{fig3}
\end{figure}

\clearpage

\begin{figure}[hbtp]
\centerline{\epsfysize20cm\epsfbox{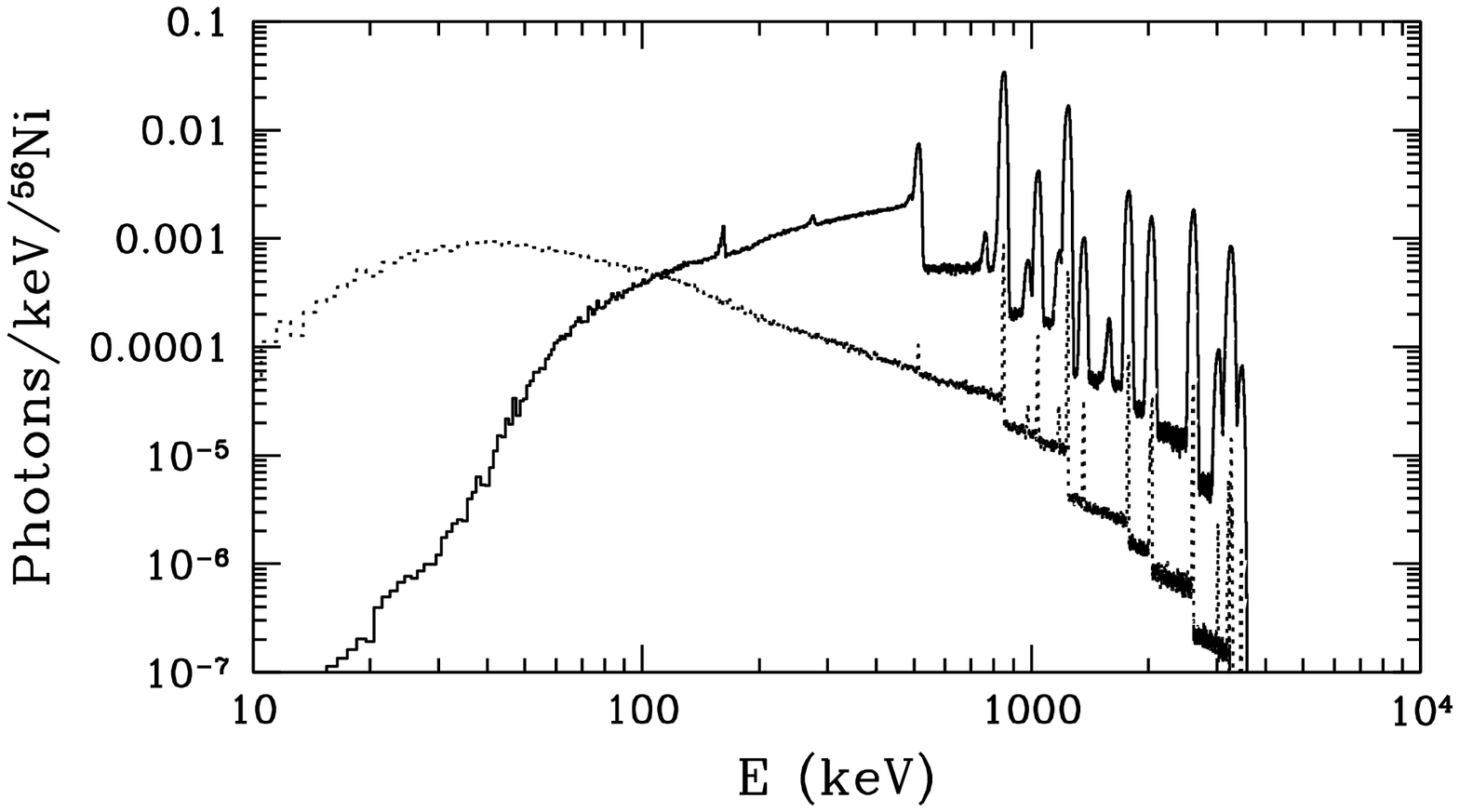}}
\nopagebreak[4]
\label{fig4}
\end{figure}

\clearpage

\begin{figure}[hbtp]
\centerline{\epsfysize20cm\epsfbox{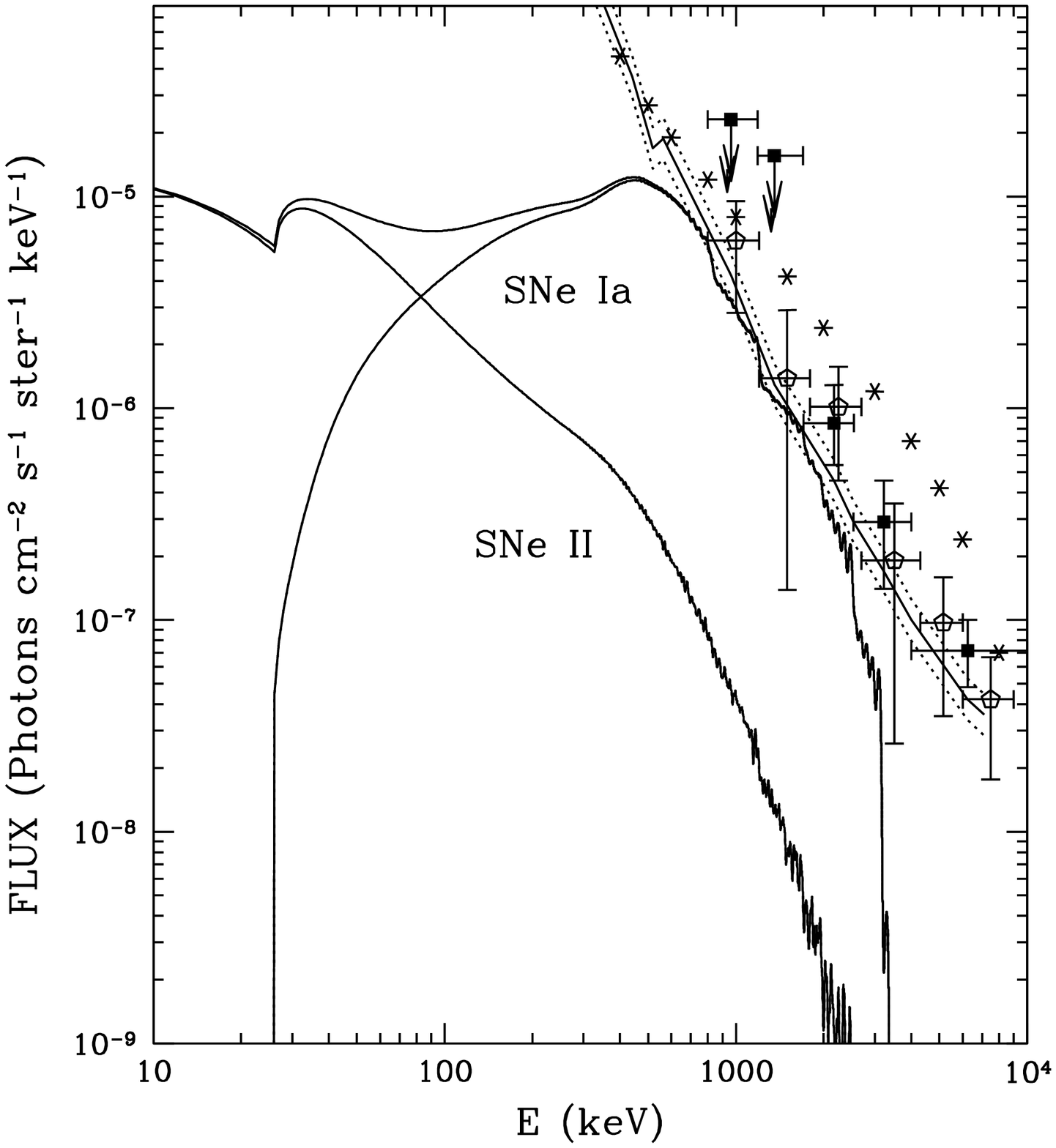}}
\nopagebreak[4]
\label{fig5}
\end{figure}

\clearpage

\begin{figure}[hbtp]
\centerline{\epsfysize20cm\epsfbox{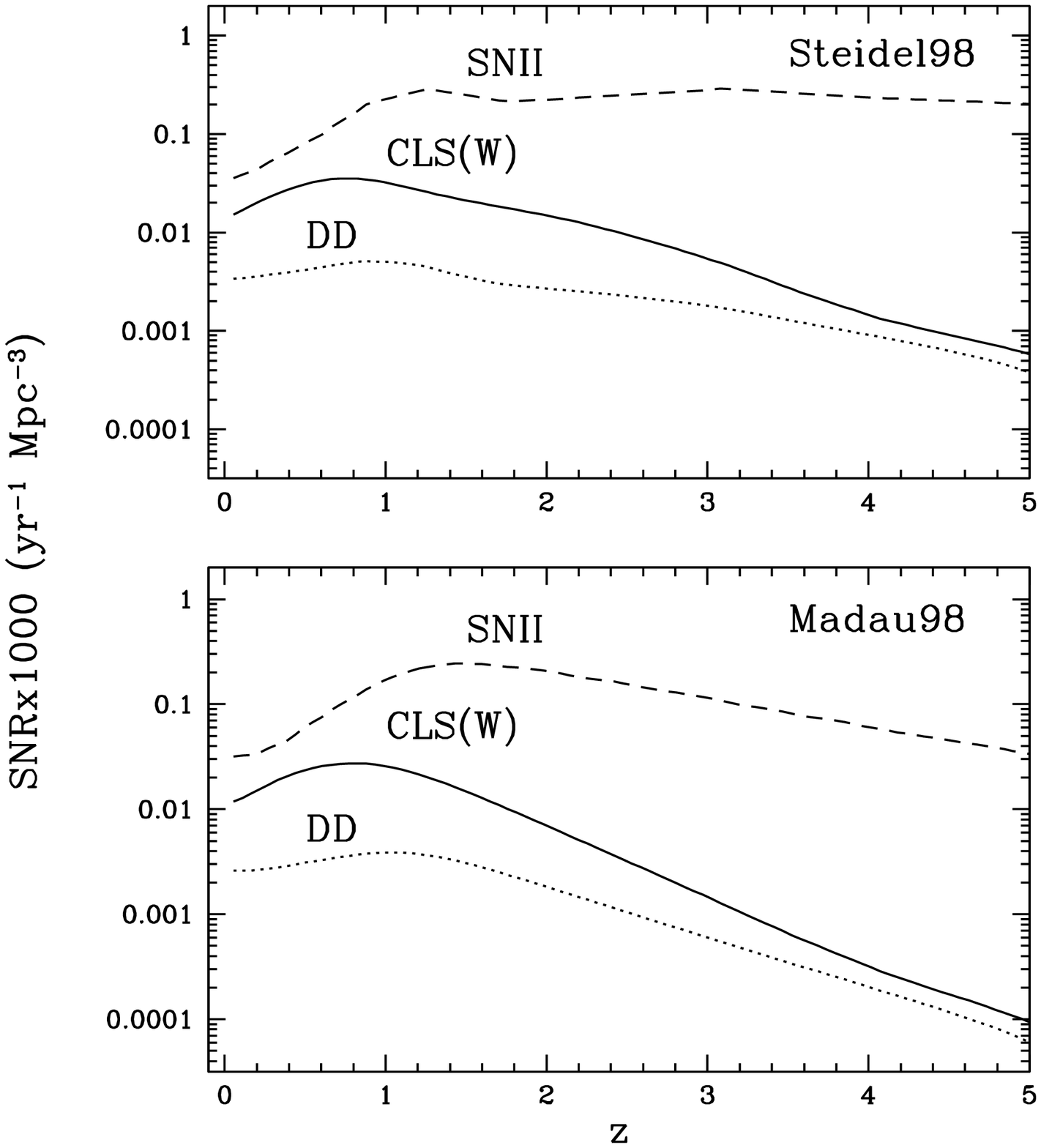}}
\nopagebreak[4]
\label{fig6}
\end{figure}

\clearpage

\begin{figure}[hbtp]
\centerline{\epsfysize20cm\epsfbox{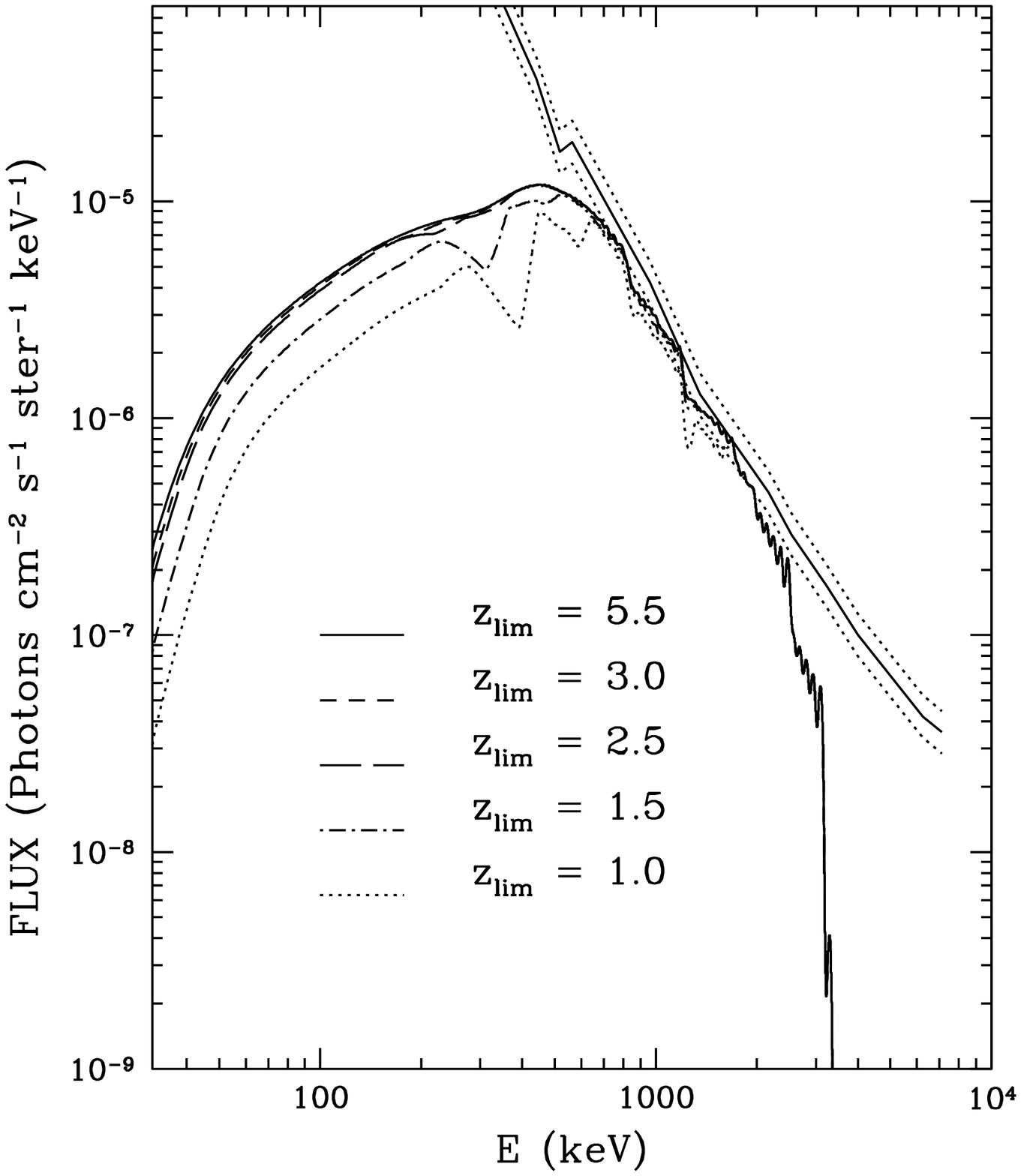}}
\nopagebreak[4]
\label{fig6}
\end{figure}

\clearpage

\begin{figure}[hbtp]
\centerline{\epsfysize20cm\epsfbox{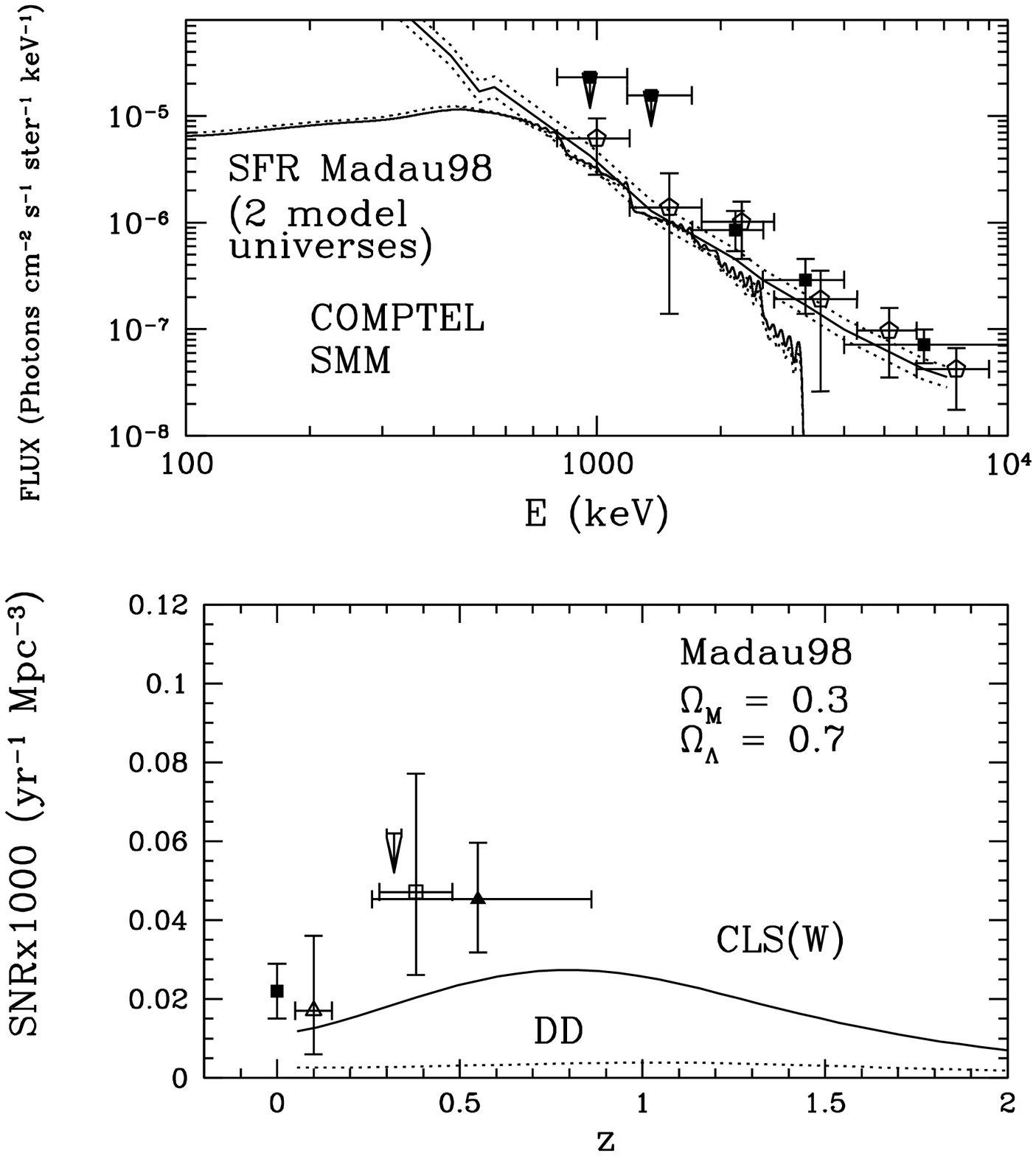}}
\nopagebreak[4]
\label{fig6}
\end{figure}

\clearpage

\begin{figure}[hbtp]
\centerline{\epsfysize20cm\epsfbox{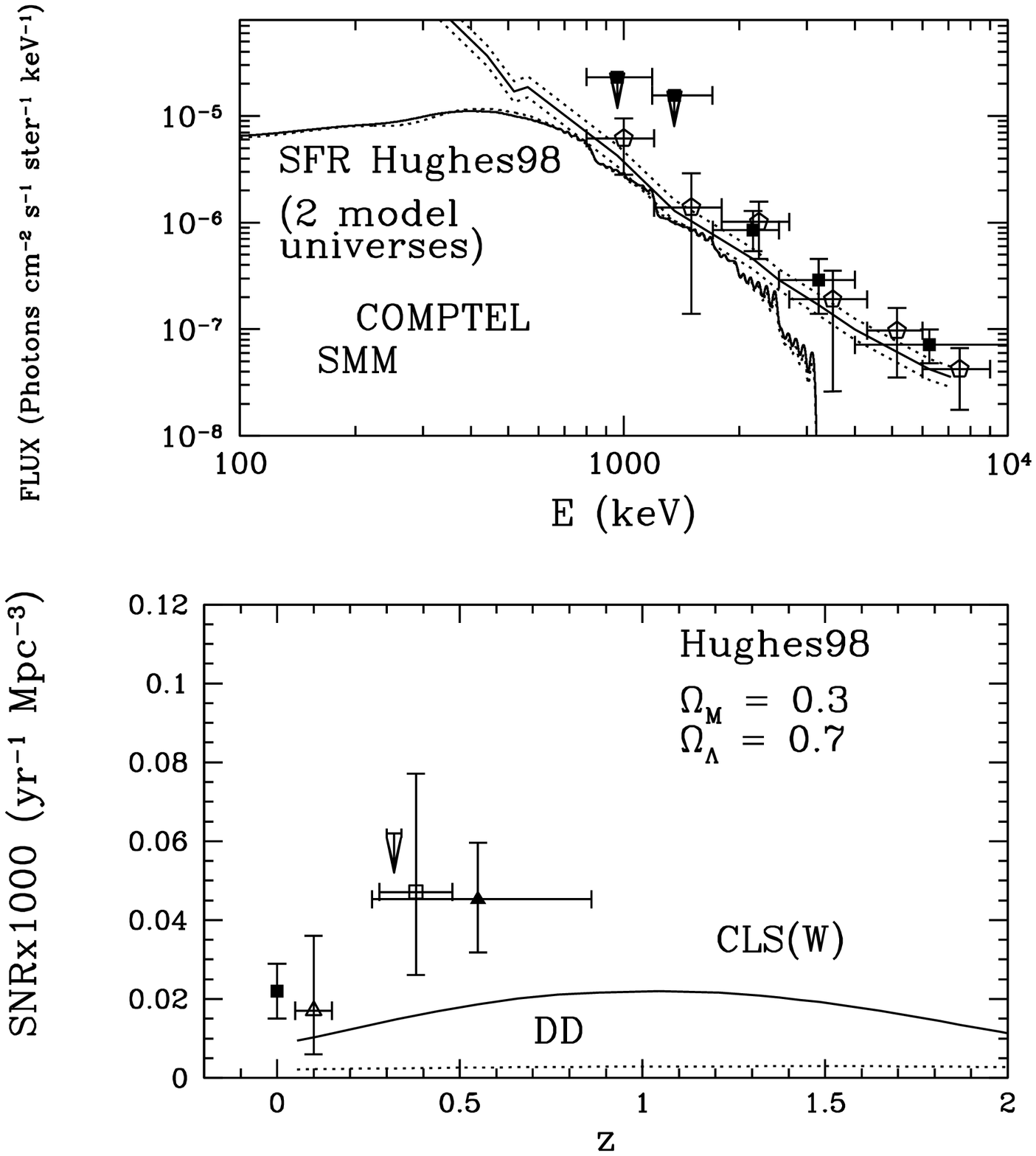}}
\nopagebreak[4]
\label{fig7}
\end{figure}

\clearpage

\begin{figure}[hbtp]
\centerline{\epsfysize20cm\epsfbox{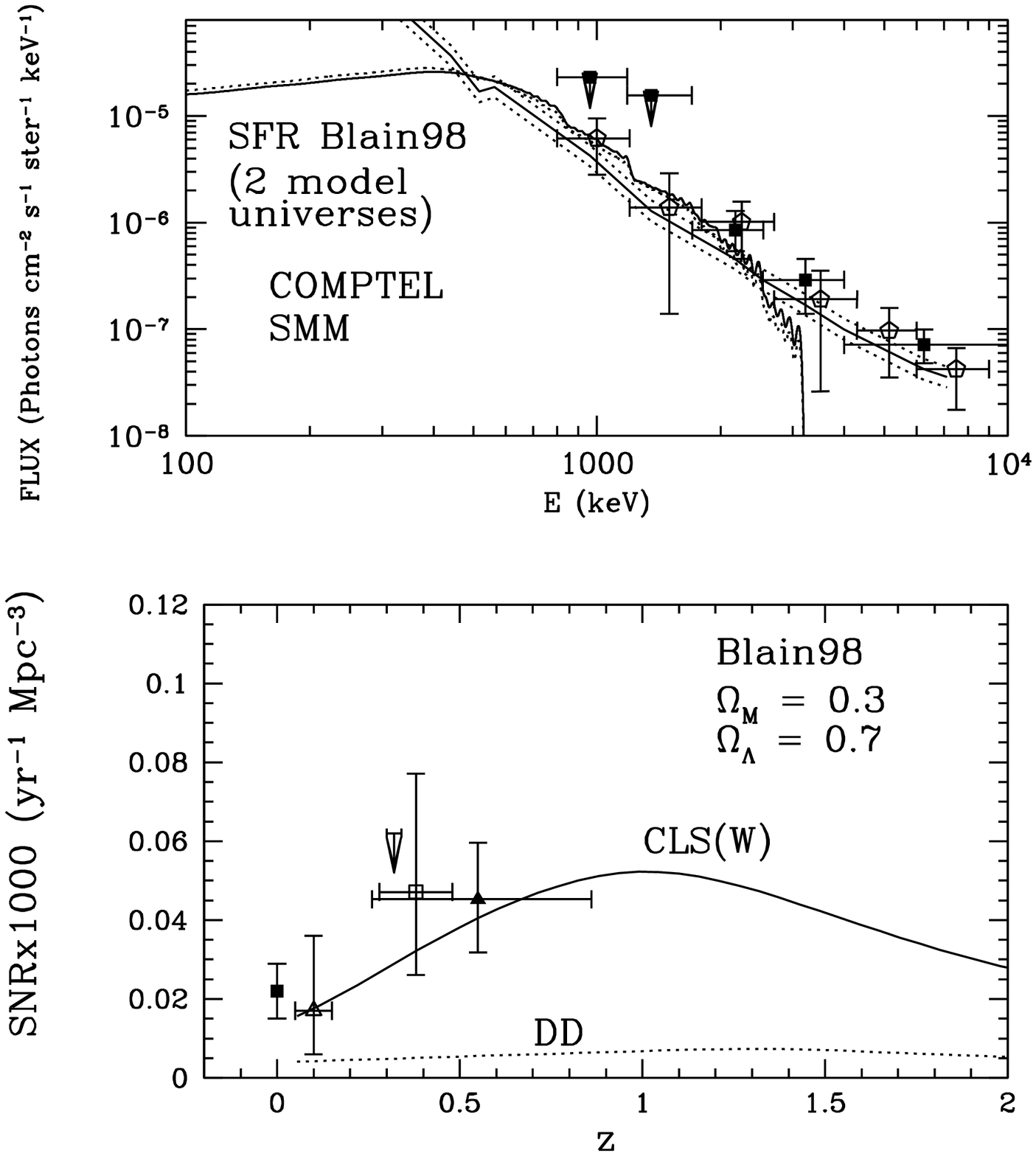}}
\nopagebreak[4]
\label{fig8}
\end{figure}

\clearpage

\begin{figure}[hbtp]
\centerline{\epsfysize20cm\epsfbox{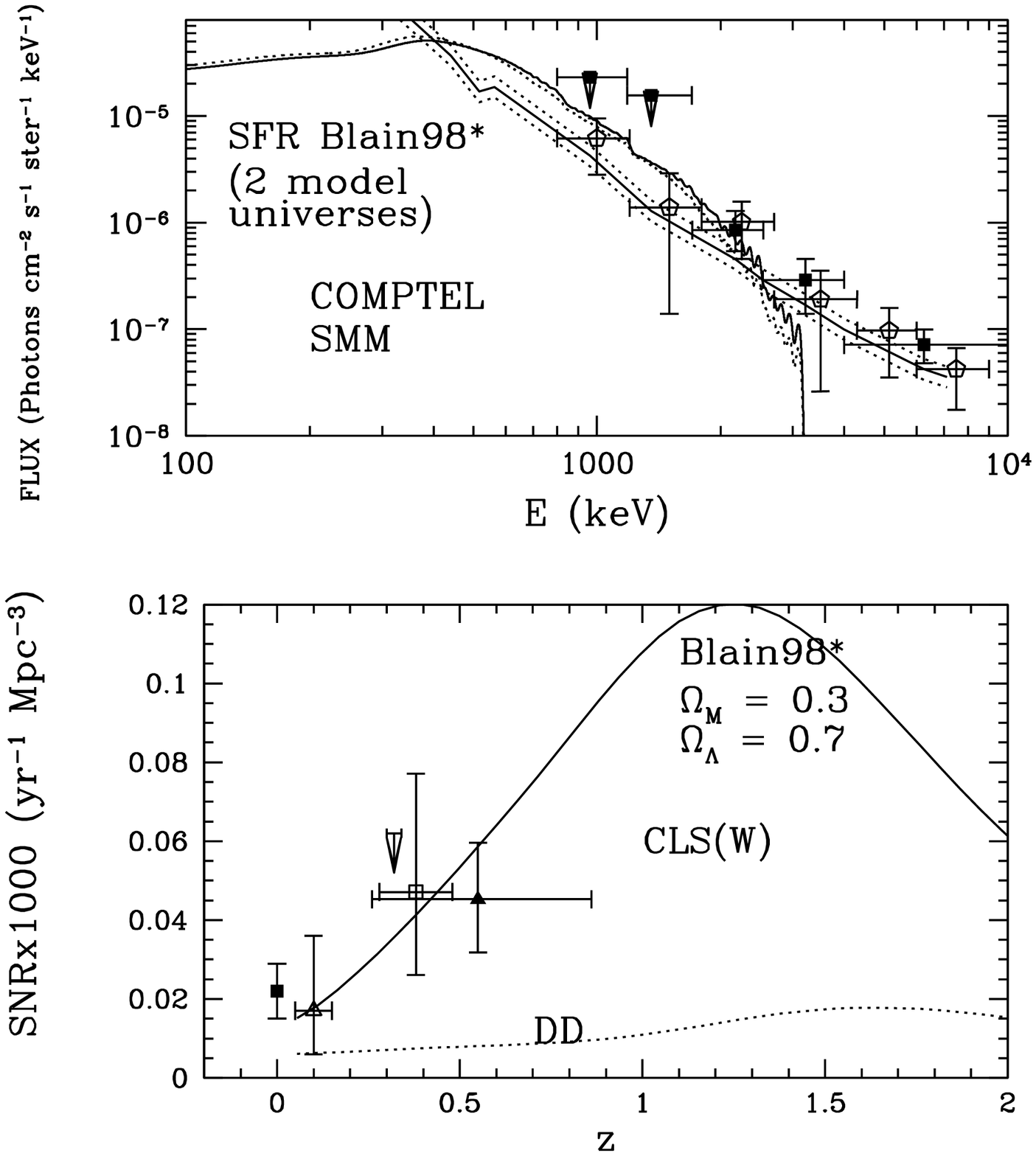}}
\nopagebreak[4]
\label{fig9}
\end{figure}

\clearpage

\begin{figure}[hbtp]
\centerline{\epsfysize20cm\epsfbox{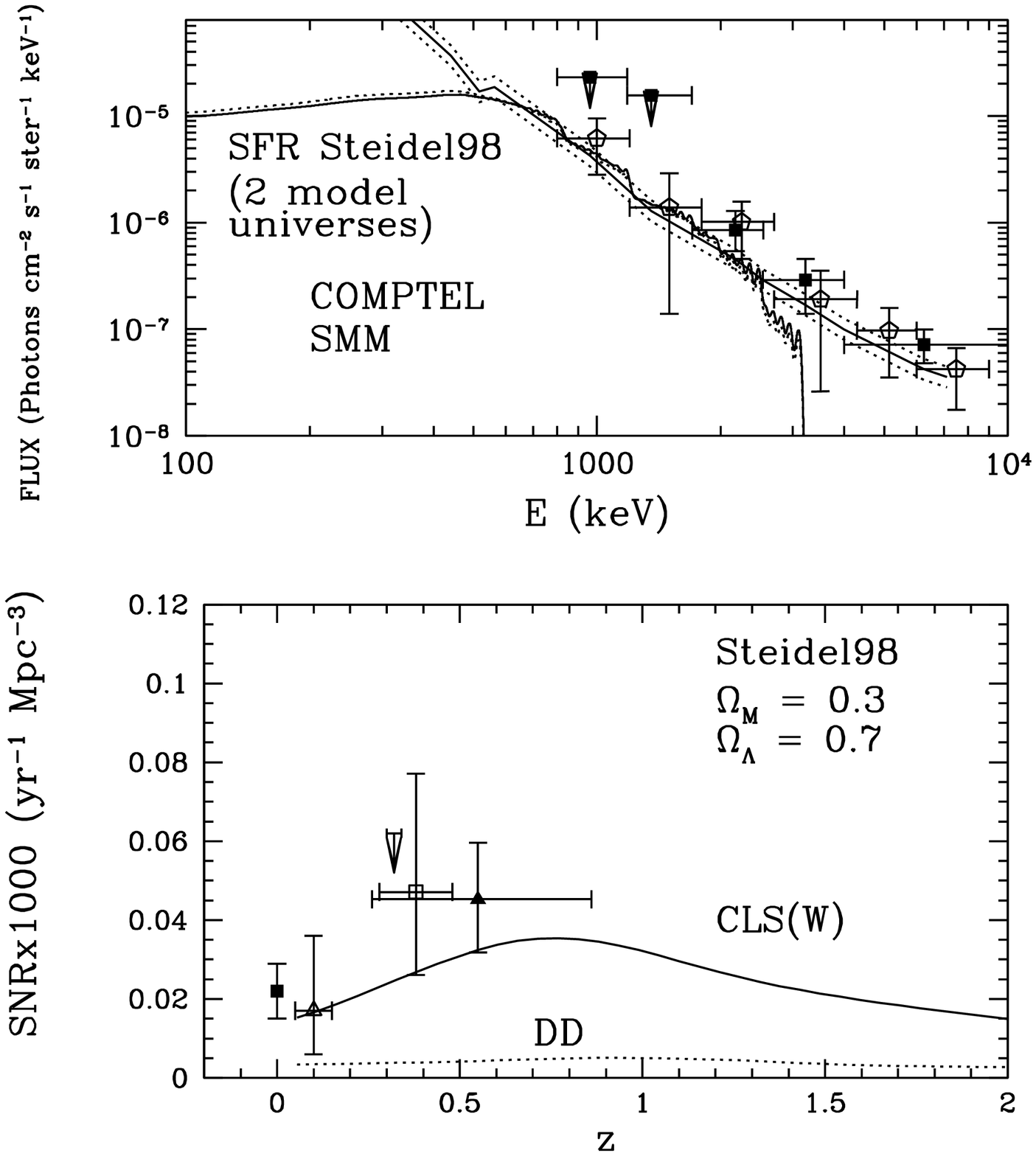}}
\label{fig10}
\end{figure}

\clearpage

\begin{figure}[hbtp]
\centerline{\epsfysize20cm\epsfbox{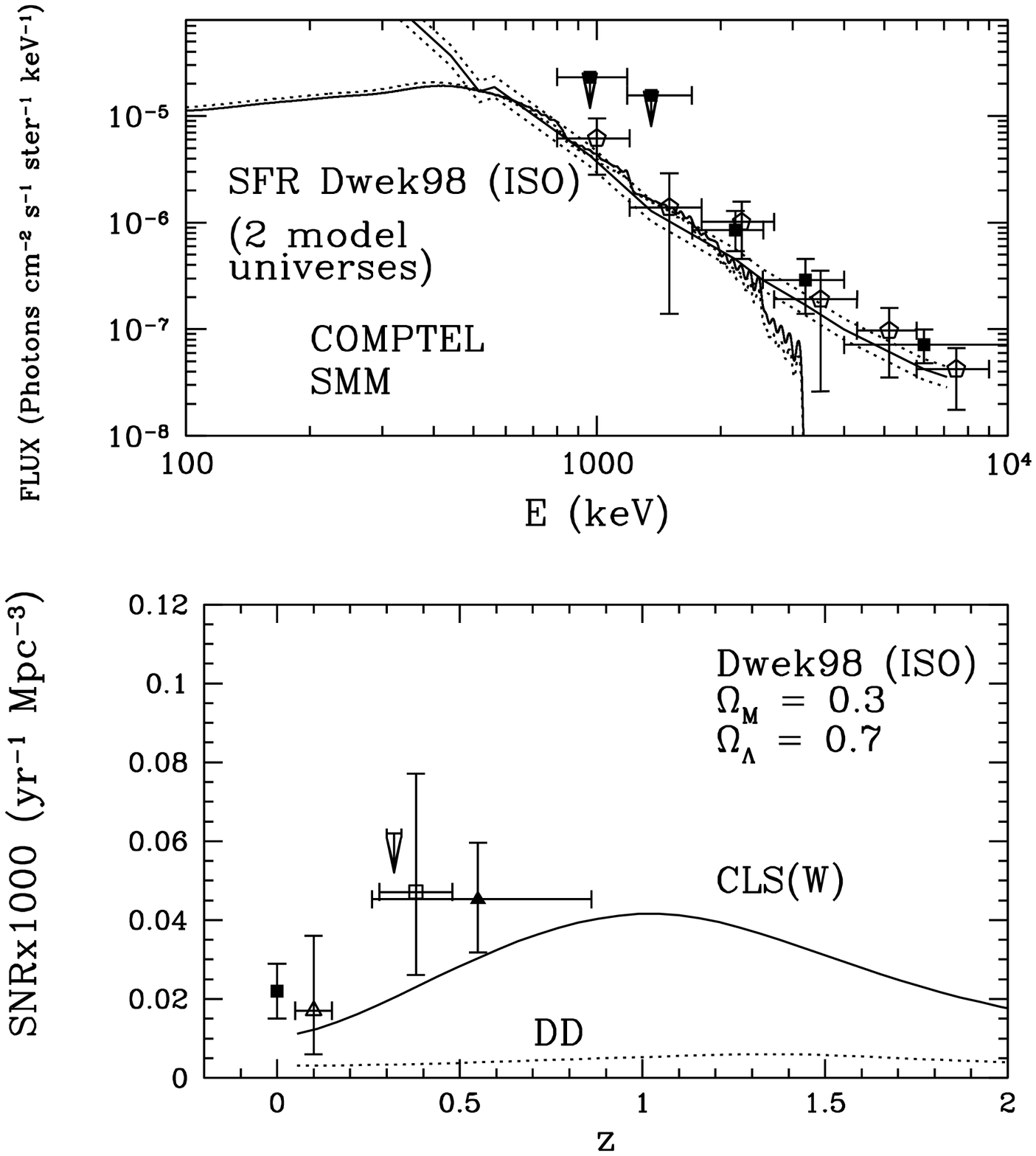}}
\label{fig11}
\end{figure}

\clearpage

\begin{figure}[hbtp]
\centerline{\epsfysize20cm\epsfbox{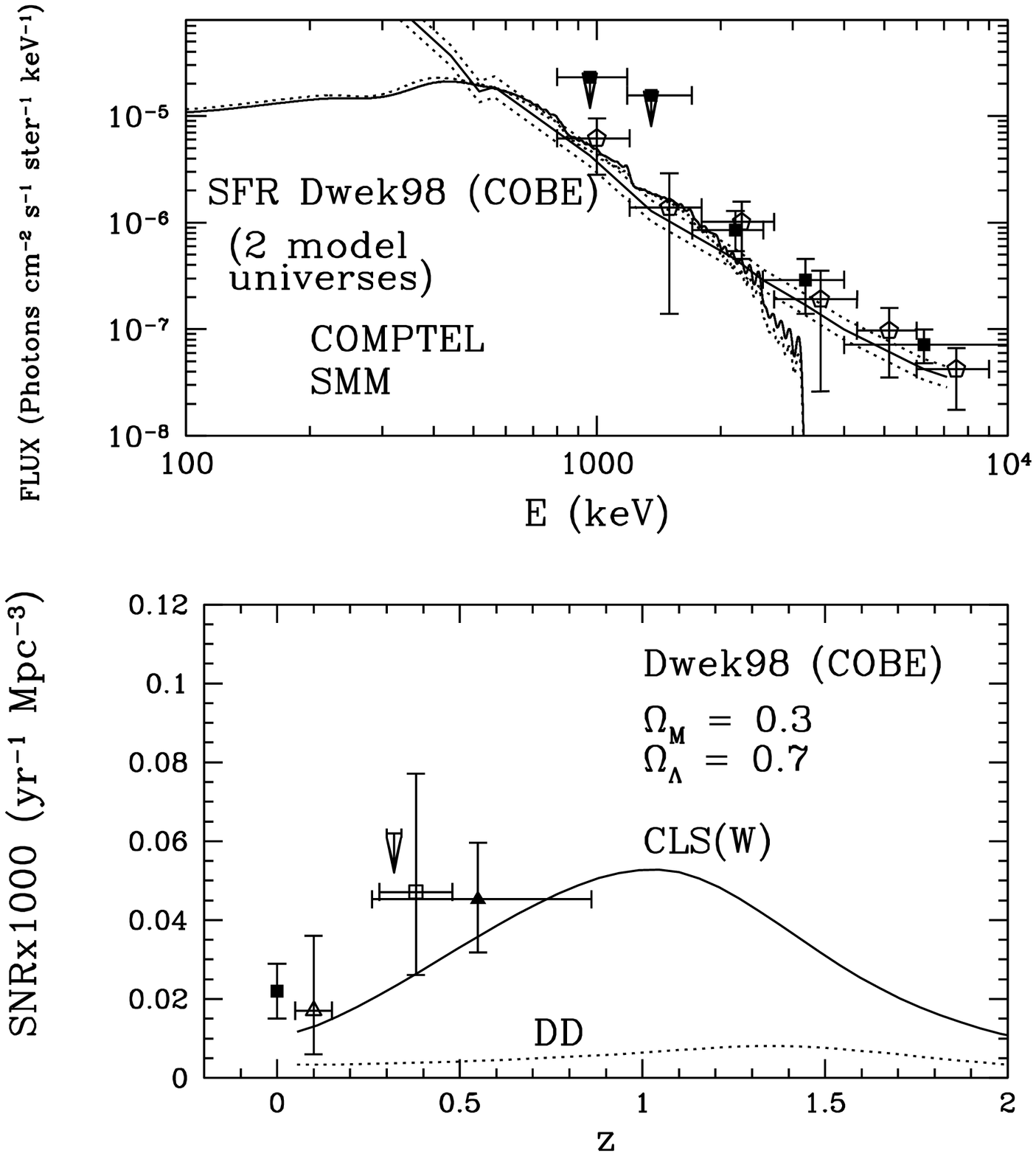}}
\label{fig12}
\end{figure}

\clearpage

\begin{figure}[hbtp]
\centerline{\epsfysize20cm\epsfbox{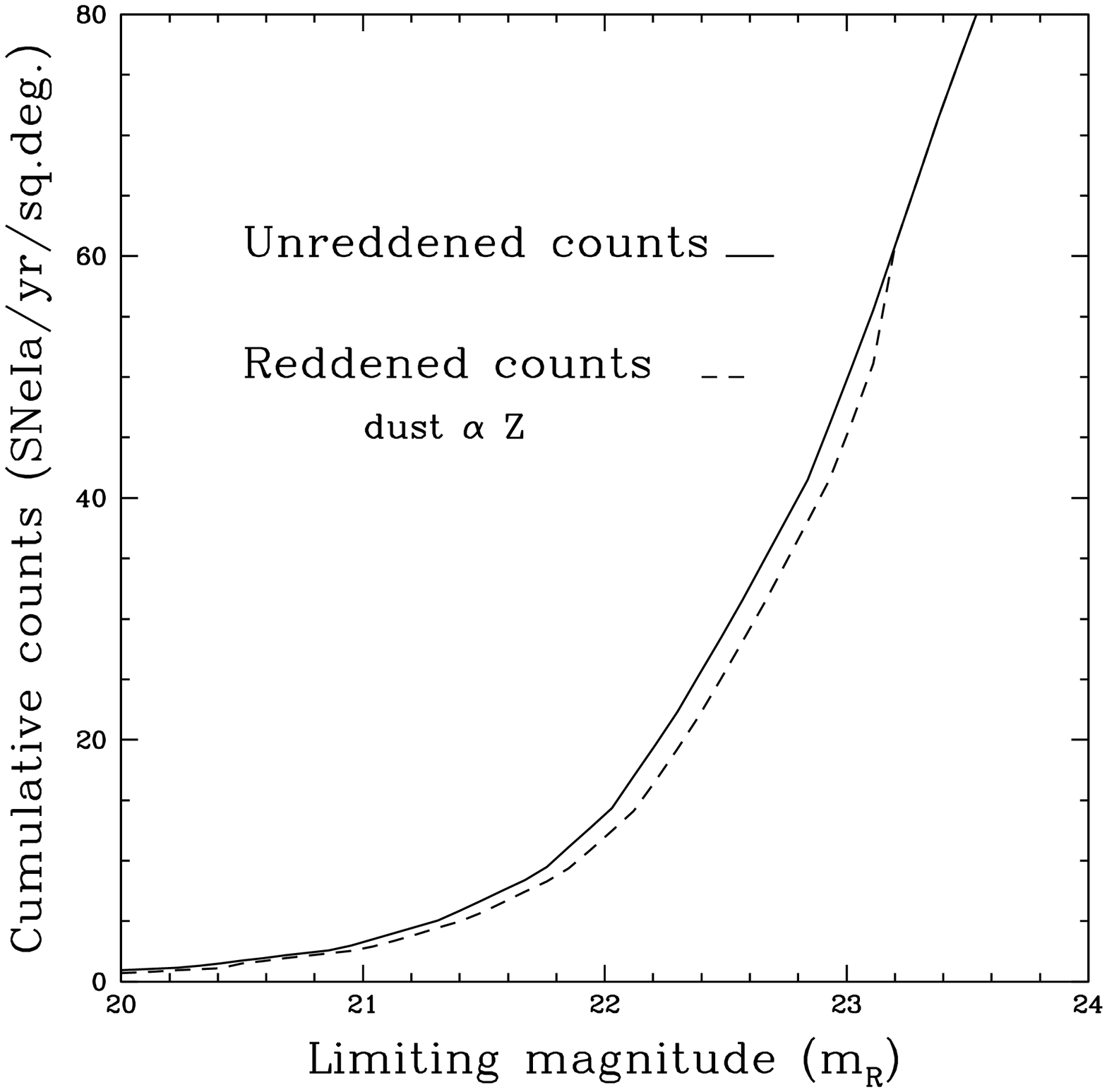}}
\label{fig13}
\end{figure}

\end{document}